\documentclass[a4paper,12pt]{article}

\newcommand{\Mg}{M_\Q }

\usepackage{mathrsfs,graphicx,rotating,amsmath,amsfonts,mathtools,booktabs,amssymb,wasysym}
\usepackage{hyperref}\usepackage{slashed}
\usepackage[nosort]{cite}
\usepackage[table,xcdraw,dvipsnames]{xcolor}
\usepackage{youngtab}
\usepackage{graphicx}
\usepackage{multirow,multicol}
\hypersetup{colorlinks,bookmarksopen,bookmarksnumbered,
linkcolor=blus,pdfstartview=FitH,urlcolor=rossos,citecolor=verde}
\allowdisplaybreaks

\newcommand{\LQCD}{\Lambda_{\rm QCD}}
\newcommand{\mio}[1]{}

 \newcommand{\med}[1]{\langle #1\rangle}

 \newcommand{\fig}[1]{~\ref{fig:#1}}
\newcommand{\sfrac}[2]{#1/#2}

\newcommand{\Q}{{\tilde{g}}}

\allowdisplaybreaks
\usepackage{multicol}
\usepackage{color}
\definecolor{rosso}{cmyk}{0,1,1,0.4}
\definecolor{rossos}{cmyk}{0,1,1,0.55}
\definecolor{rossoc}{cmyk}{0,1,1,0.2}
\definecolor{blu}{cmyk}{1,1,0,0.3}
\definecolor{blus}{cmyk}{1,1,0,0.6}
\definecolor{bluc}{cmyk}{1,1,0,0.1}
\definecolor{verde}{cmyk}{0.92,0,0.59,0.25}
\definecolor{verdec}{cmyk}{0.92,0,0.59,0.15}
\definecolor{verdes}{cmyk}{0.92,0,0.59,0.4}

\newcommand{\eq}[1]{~{\rm (\ref{eq:#1})}}

\newcommand{\s}{\,{\rm s}}

\newcommand{\MeV}{\,{\rm MeV}}
\newcommand{\GeV}{\,{\rm GeV}}
\newcommand{\TeV}{\,{\rm TeV}}
\newcommand{\PeV}{\,{\rm PeV}}

\newcommand{\Tr}{\,{\rm Tr}}

\def\circa#1{\,\raise.3ex\hbox{$#1$\kern-.75em\lower1ex\hbox{$\sim$}}\,}

\newcommand{\beq}{\begin{equation}}
\newcommand{\eeq}{\end{equation}}

\newcommand{\bea}{\begin{eqnarray}}
\newcommand{\eea}{\end{eqnarray}}
\newcommand{\be}{\begin{equation}}
\newcommand{\ee}{\end{equation}}
\font\tenrsfs=rsfs10 at 12pt
\font\sevenrsfs=rsfs7
\font\fiversfs=rsfs5
\newfam\rsfsfam
\textfont\rsfsfam=\tenrsfs
\scriptfont\rsfsfam=\sevenrsfs
\scriptscriptfont\rsfsfam=\fiversfs

\newsavebox\MBox

\newcommand{\LDC}{\Lambda_{\rm QCD}}

\newcommand{\alf}{\alpha_{\rm eff}}

\newcommand{\SU}{\,{\rm SU}}

\def\circa#1{\,\raise.3ex\hbox{$#1$\kern-.75em\lower1ex\hbox{$\sim$}}\,}
\makeatletter

\font\ital=cmu10 

\def\hhref#1{\href{http://arxiv.org/abs/#1}{arXiv:#1}}
\usepackage{xstring} 
\newcommand{\hhrefq}[1]{\IfSubStr{#1}{:}{\href{http://inspirehep.net/search?ln=en&ln=en&p=#1&of=hb&action_search=Search&sf=&so=d&rm=&rg=25&sc=0}{InSpires:#1}}{\hhref{#1}}}

\def\art{\@ifnextchar[{\eart}{\oart}}
\def\eart[#1]#2#3#4#5#6{{\rm #2}, {\em #3 \bf #4} {\rm (#6) #5} ({\em #1})}
\def\article{\@ifnextchar[{\earticle}{\oarticle}}
\def\oarticle#1#2#3#4#5#6{{\rm #1}, {\ital ``#6''}, {\rm #2 #3 (#5) #4}}
\def\earticle[#1]#2#3#4#5#6#7{{\rm #2}, {\ital ``#7''}, {\rm #3 #4 (#6) #5}  [\hhrefq{#1}]}
\def\hepart[#1]#2{{\rm #2, \sl#1}}
\def\heparticle[#1]#2#3{#2, {\ital ``#3''} [\hhrefq{#1}]}
\newcommand{\doi}[1]{\href{http://dx.doi.org/#1}{[link]}}

\renewenvironment{thebibliography}[1]
     {\begin{multicols}{2}[\section*{\refname}]%
      \@mkboth{\MakeUppercase\refname}{\MakeUppercase\refname}%
      \list{\@biblabel{\@arabic\c@enumiv}}%
           {\settowidth\labelwidth{\@biblabel{#1}}%
            \leftmargin\labelwidth
            \advance\leftmargin\labelsep
            \@openbib@code
            \usecounter{enumiv}%
            \let\p@enumiv\@empty
            \renewcommand\theenumiv{\@arabic\c@enumiv}}%
      \sloppy
      \clubpenalty4000
      \@clubpenalty \clubpenalty
      \widowpenalty4000%
      \sfcode`\.\@m}
     {\def\@noitemerr
       {\@latex@warning{Empty `thebibliography' environment}}%
      \endlist\end{multicols}}

\newcounter{alphaequation}[equation]
\def\thealphaequation{\theequation\hbox to
0.6em{\hfil\alph{alphaequation}\hfil}}
\def\eqnsystem#1{
\def\@eqnnum{{\rm (\thealphaequation)}}
\def\@@eqncr{\let\@tempa\relax \ifcase\@eqcnt \def\@tempa{& & &} \or
  \def\@tempa{& &}\or \def\@tempa{&}\fi\@tempa
  \if@eqnsw\@eqnnum\refstepcounter{alphaequation}\fi
\global\@eqnswtrue\global\@eqcnt=0\cr}
\refstepcounter{equation} \let\@currentlabel\theequation \def\@tempb{#1}
\ifx\@tempb\empty\else\label{#1}\fi
\refstepcounter{alphaequation}
\let\@currentlabel\thealphaequation
\global\@eqnswtrue\global\@eqcnt=0 \tabskip\@centering\let\\=\@eqncr
$$\halign to \displaywidth\bgroup \@eqnsel\hskip\@centering
$\displaystyle\tabskip\z@{##}$&\global\@eqcnt\@ne
\hskip2\arraycolsep\hfil${##}$\hfil& \global\@eqcnt\tw@\hskip2\arraycolsep
$\displaystyle\tabskip\z@{##}$\hfil
\tabskip\@centering&\llap{##}\tabskip\z@\cr}
\def\endeqnsystem{\@@eqncr\egroup$$\global\@ignoretrue} \makeatother

\newcommand{\xxx}[1]{{\color{red}[\bf #1]}}

\definecolor{Gray}{gray}{0.95}

\begin{document}
{\hfill CP3-Origins-2018-034 DNRF90\hfill}
\vspace{1cm}

\begin{center}
{\Large\LARGE \bf \color{rossos}
Cosmological Abundance of Colored Relics}\\[0.8cm]
{\bf Christian Gross}$^{a,b}$,
{\bf Andrea Mitridate$^{b,c}$,\\
Michele Redi$^{d,e}$,
 Juri Smirnov$^{f}$, 
Alessandro Strumia$^{a}$}  
\\[7mm]
{\it $^a$ Dipartimento di Fisica dell'Universit{\`a} di Pisa, Italy}\\
{\it $^b$ INFN, Sezione di Pisa, Italy}\\
{\it $^c$ Scuola Normale Superiore, Pisa, Italy}\\
{\it $^d$ INFN sezione di Firenze, Via G. Sansone 1; I-59100 Sesto F.no, Italy}\\ 
{\it $^e$ Department of Physics and Astronomy, University of Florence, Italy}\\
{\it $^f$ $\text{CP}^3$-Origins and DIAS, University of Southern Denmark, Odense, Denmark}\\

\vspace{1.0cm}

{\large\bf\color{blus} Abstract}
\begin{quote}\large
The relic cosmological abundance of stable or long-lived neutral
colored particles gets reduced by about a few orders of magnitude
by annihilations that occur after QCD confinement.
We compute the abundance and the cosmological bounds on relic gluinos.
The same post-confinement
effect strongly enhances co-annihilations with a lighter Dark Matter particle, provided
that their mass difference is below a few GeV.
Charged colored particles (such as stops)
can instead form baryons, which can be (quasi)stable in some models.

\end{quote}
\thispagestyle{empty}
\end{center}

\newpage

\tableofcontents

\setcounter{footnote}{0}

\section{Introduction}
Extensions of the Standard Model (SM) sometimes predict (quasi)stable colored particles.
We show that, due to non perturbative QCD effects, their relic abundance is significantly lower than previously expected, correspondingly reducing the phenomenological constraints.

Weak-scale supersymmetry has been considered the
most motivated extension of the SM, as it allows to control quadratically divergent corrections to the Higgs mass
keeping them  naturally small;
improves  the prediction for the gauge couplings
in SU(5) unification; provides Dark Matter (DM) candidates.
 The plausibility of the naturalness goal is now endangered  by
the lack of any new physics in LEP~\cite{hep-ph/9811386} and LHC data~\cite{1101.2195}.
Furthermore the Higgs mass is larger than what predicted by the MSSM with weak-scale sparticles.

Split SuperSymmetry~\cite{hep-th/0405159,hep-ph/0409232} 
(where the new supersymmetric fermions are much lighter than the new supersymmetric scalars)
abandoned the naturalness goal, retaining the two other good features,
allowing to fit the Higgs mass~\cite{1108.6077,1407.4081},
and relaxing the possible supersymmetric flavour problem caused by weak-scale sfermions.
If sfermions are very heavy the light gauginos can become long-lived, giving peculiar signatures at colliders and potential cosmological problems.
These were explored in~\cite{Mina}, where the relic gluino abundance (before late gluino decay in neutralino and colored SM particles) was computed including perturbative gluino annihilations 
at $T \sim M_3$ and
arguing that one can neglect non-perturbative effects arising after confinement at $T \sim \LQCD$.
Such effects reduce the relic gluino abundance by a few orders of magnitude~\cite{1801.01135}, thereby weakening  cosmological bounds. 

The relevance of confinement effects has been estimated  in~\cite{hep-ph/0611322} 
in the case of colored charged particles.
Unlike in the case of the neutral gluino, QCD bound states of 
charged particles can be formed or broken by emitting or absorbing photons.
We will consider the case of (quasi)stable stop $\tilde t$.

In section~\ref{gluino} we compute the thermal relic abundance
of (quasi)stable gluinos and
in section~\ref{pheno} we reconsider the cosmological bounds and discuss the
associated phenomenology.  
Conclusions are given in section~\ref{concl}.

%
%

\section{Relic gluinos}\label{gluino}
We consider a Majorana fermion in the adjoint of SU(3). In supersymmetric models this is known as gluino and denoted as $\tilde g$.
The gluino can be stable if it is the lightest supersymmetric particle.
Otherwise it can decay via squark exchange into a quark, an antiquark and a neutralino or chargino, or radiatively to a gluon and a neutralino, with quarks and squarks in the loop.
The resulting lifetime is long if sfermions have a much heavier mass $m_{\rm SUSY}$~\cite{Toharia:2005gm,Gambino:2005eh}:
\be\label{eq:taug}
\tau_{\tilde g}=\frac{4 \ \textrm{sec}}{N} \left(\frac{m_{\rm SUSY}}{10^9 \GeV}\right)^4 \left(\frac{\TeV}{\Mg}\right)^5 \,,
\ee
where $N$ is an order-one function~\cite{Gambino:2005eh}.
A stable or long lived gluino is probed and constrained by cosmology.

 \begin{figure}[t]
\begin{center}
\includegraphics[width=.81\textwidth]{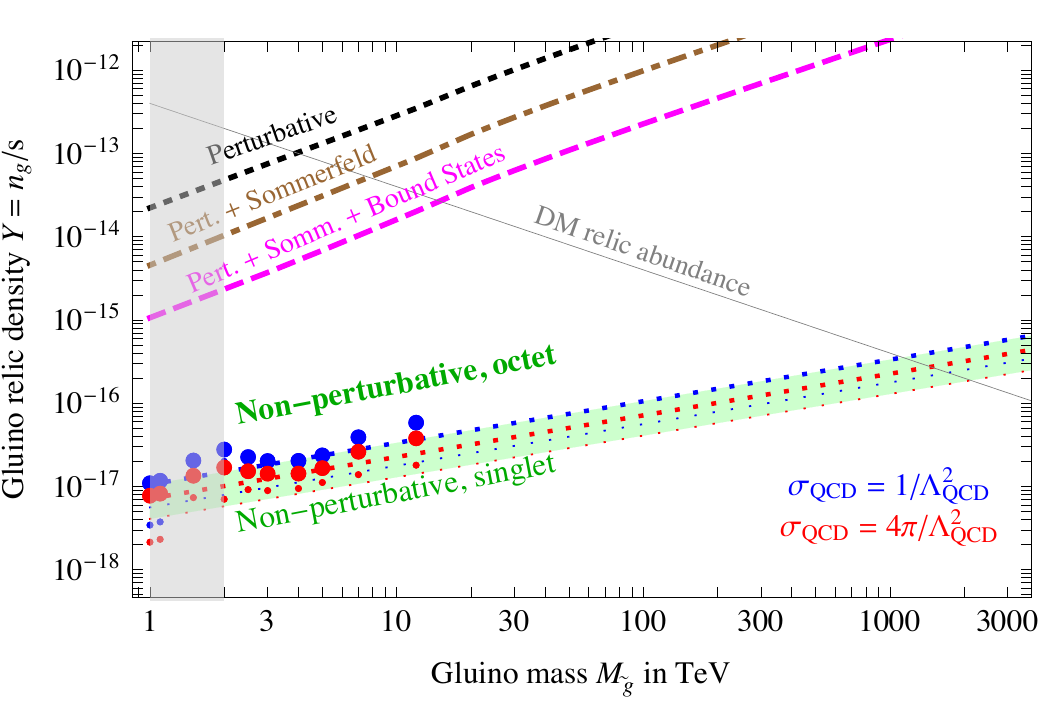}\qquad
\caption{\em \label{fig:Omega} Predicted gluino abundance. 
Relic stable gluinos exceed the DM density if $M_\Q \circa{>} \PeV$.
The bands show the non-perturbative analytic result for $\sigma_{\rm QCD} =1/\LQCD^2$ (blue) and
$\sigma_{\rm QCD} =4\pi/\LQCD^2$ (red). 
The thin (thick) lines assume that only singlet bound states (octet bound states too) can form with QCD size;
similarly, the small (large)  dots show our numerical computation for some values of the gluino mass.}
\end{center}
\end{figure}

\subsection{Computing the relic gluino abundance}\label{relicgluino}
Fig.~\ref{fig:Omega} shows our result for the gluino relic abundance, before their possible slow decays.
This is computed as follows.
The upper curves show the relic abundance after a first decoupling at $T \sim \Mg/25$,
as computed in various approximations:
\begin{enumerate}
\item at tree level in the perturbative expansion;
\item  taking into account  Sommerfeld corrections 
the  $s$-wave annihilation cross-section is (eq.~(2.24) of~\cite{1402.6287},
where the Sommerfeld $S$ factors are defined) 
\begin{equation}
\sigma_{\rm ann} v_{\rm rel} =\frac {27} {32}\sigma_0 \left[\frac 1 6 S_{3}+\frac 1 3 S_{ 3 /2}+ \frac 1 2  S_{-1}\right]+ 
 \frac {9} {8} \sigma_0  S_{3/2},\qquad \sigma_0=\frac{\pi \alpha_3^2}{M_\Q ^2}
\end{equation}
\item  taking into account also a related effect: formation of bound states~\cite{1702.01141}.
\end{enumerate}
These effects reduce by about 1 order of magnitude the gluino abundance,
controlled by the Boltzmann equation
\begin{align} \label{eq:eff}
&\frac{Hz}{s} \frac{dY_{\Q}}{dz} = -
 \med{\sigma_{\rm ann} v_{\rm rel}}  (Y_\Q^2  -Y_\Q^{\rm eq2})
 \end{align}
where $z = M_\Q/T$, $Y_{\Q} = n_{\Q}/s$, $s$ is the entropy density at temperature $T$;
$H(T)$ is the Hubble constant.

If $\tau_{\tilde g}< M_{\rm Pl}/\LQCD^2 \sim\mu{\rm sec}$ gluinos decay before the QCD phase transition
leaving no cosmological effects.
Otherwise gluinos recouple as the temperature approaches the QCD scale, and their relic abundance is
determined by a  re-decoupling at temperatures mildly below the QCD phase transition.
At this point gluinos have formed $\tilde g g$ and/or $\tilde g q \bar q'$ hadrons which scatter with
large cross sections $\sigma_{\rm QCD} = c/\LQCD^2$ where $c \sim 1$, making about $M_{\rm Pl}/\LQCD \sim 10^{19}$
scatterings in a Hubble time.  
For comparison,
the proton-proton elastic scattering cross section at low energy is known to be
$\sigma_{\rm el}  \approx 100 \, \rm mb$, corresponding to $c\approx 23$.

Although gluinos are much rarer than gluons and quarks, occasionally, two gluino hadrons
meet forming a $\tilde g\tilde g$ bound state.
Classically such state has angular momentum
$\ell \approx \mu v_{\rm rel} b$ where $b\approx 1/\LQCD$ is the impact parameter;
$\mu\simeq \Mg/2$ is the reduced mass; 
$v_{\rm rel}\sim ({T/\Mg})^{1/2}$ is the relative velocity.
Thereby $\ell \sim (\Mg T)^{1/2}/\LQCD$,  is large
for $\Mg \gg \LQCD \circa{>} T$. 
The quantum-mechanical total QCD cross section for forming $\tilde g\tilde g$ bound states
is large because many partial waves contribute.
This can be parameterized defining the maximal angular momentum as
$\ell_{\rm max}  \equiv   \sqrt{c/2\pi} M_\Q v_{\rm rel}/\LQCD $ 
obtaining (see e.g.~\cite{Griest:1989wd})
\beq\label{eq:sigmaQCD}
\sigma_{\rm QCD} = \sum_{\ell =0}^{\ell_{\rm max}}  \sigma_\ell \simeq
 \frac{ c }{\LQCD^2},\qquad
\sigma_\ell =4\pi   \frac{ 2\ell +1 }{ M_\Q^2 v_{\rm rel}^2} \sin^2\delta_\ell.
\eeq
where the phase shifts average to $\langle \sin^2\delta_\ell  \rangle\simeq 1/2$.
This expectation is consistent with numerical results in toy calculable models~\cite{1802.07720}.

The cross section
relevant for reducing the gluino abundance is not $\sigma_{\rm QCD}$,
but the smaller cross section $\sigma_{\rm ann}$ for forming 
$\tilde g\tilde g$ states which annihilate into SM particles before being broken.
Assuming that a $\tilde g\tilde g$ with angular momentum $\ell$ and energy $\sim T$
annihilates before being broken with probability $\wp_\ell(T)$, one has\footnote{This intuitive picture can be formally justified writing a network of Boltzmann equations,
one for each bound state $I$ with different $\ell$ and $n$.
Such equations contain the formation rates $\gamma_I$, the
thermally averaged breaking rates $\Gamma_I^{\rm break}$, 
the annihilation rates $\Gamma_I^{\rm ann}$,
the decay rates among the states $\Gamma_{IJ}$.
This is unpractical, given that hundreds of states play a relevant role.
To get some understanding, we consider a toy system where only one state 1 can be produced,
and only one state 3 can annihilate. 
The state 1 can decay to state 2, which can decay to state 3.
Then, assuming that the rates are faster than the Hubble rate, 
one can reduce the network of Boltzmann equations \cite{1702.01141} to the single Boltzmann equation eq.\eq{eff}
for the total gluino density, controlled by an effective annihilation  rate equal to $\wp \gamma_1$ where
\beq \label{eq:BR} \wp = \hbox{BR}_{12}\hbox{BR}_{23} ,\qquad
\hbox{BR}_{12} =\frac{\Gamma_{12}}{\Gamma_{12}+\Gamma_{1}^{\rm break}},\qquad
\hbox{BR}_{23} =\frac{\Gamma_{23}}{\Gamma_{23}+\Gamma_{2}^{\rm break}+
\hbox{BR}_{12}\Gamma_{1}^{\rm break} 
}\eeq
where the last term takes into account that 2 can upscatter to 1.
We see that $\wp$ does not depend on $\Gamma_3^{\rm ann}$
and has the expected physical meaning.
In view of QCD uncertainties  we cannot compute all order unity factors, 
such that it is appropriate to employ the simpler intuitive picture.}
\beq \label{eq:ellcross}
\sigma_{\rm ann} = \sum_{\ell =0}^{\ell_{\rm max}}  \sigma_\ell\wp_\ell.
\eeq
%
A large cross section needs large $\ell$, but $\wp_\ell$ can be small at large $\ell$.
We compute $\wp_\ell$ as  the probability that the $\Q\Q$ bound state
radiates an energy big enough to become unbreakable (bigger than $\approx T$)
before the next collision, after a time 
$\Delta t \sim 1/n_\pi v_\pi \sigma_{\rm QCD} $.
In such a case it becomes unbreakable and keeps radiating until  $\tilde g\tilde g$ annihilate.

The key quantity to be computed is thereby the power radiated by the relevant bound states
which have $n,\ell\gg1$.
In the abelian case, this is well approximated by its classical limit: Larmor radiation.
Having assumed neutral constituents, we can neglect photon radiation.
Similarly, gravitational radiation has cosmologically negligible rates $\Gamma_{\rm grav}\sim E_B^3/M_{\rm Pl}^2$.
The dominant radiation mechanism is gluon radiation, which differs from
abelian radiation because gluons are charged under QCD.
This makes a difference when (as in our case) particles are accelerated because of the strong force itself.
While a photon can be soft and its emission leaves the bound state roughly unchanged,
an emitted gluon has its own QCD potential energy, and its emission changes the QCD potential
among gluinos by an order one amount (in particular, a singlet bound state becomes octet).
As the classical limit of gluon emission is not known, we apply the quantum formul\ae{}.



We need to compute the power radiated by highly excited bound states, with sizes of order $1/\LQCD$.
Smaller bound states can be approximated by the Coulomb-like non-relativistic limit of the QCD potential,
and can have various color configurations, in particular singlets and octets.
At large distances, they appear as color singlets because they are surrounded 
by a soft gluon cloud  at distance of order $1/\LQCD$, which acts as a spectator
when computing their inner behaviour.
In the opposite limit, states larger than $1/\LQCD$ can only be  color-singlet hadrons.
For our purpose what is needed are QCD-size bound states which are the most challenging, as
confinement effects are starting to be relevant.
We will estimate their effect into two opposite limits:
\begin{itemize}
\item[8)] assuming that color octet bound states are relevant, such that 
radiation is dominated by single-gluon emission 
(pion emission after hadronization) into singlet states.
This is computed in section~\ref{8}.

\item[1)] assuming that only color singlets exists, such that radiation is dominated by
color-singlet double-gluon emission (pion emission after hadronization) among singlets.
This is computed in section~\ref{1}.
\end{itemize}
While the two cases are analytically very different (e.g.\ different powers of the strong coupling),
QCD is relatively strongly coupled so that the numerical final results in the two limiting cases will be similar.

 \begin{figure}[t]
\begin{center}
\includegraphics[width=.57\textwidth]{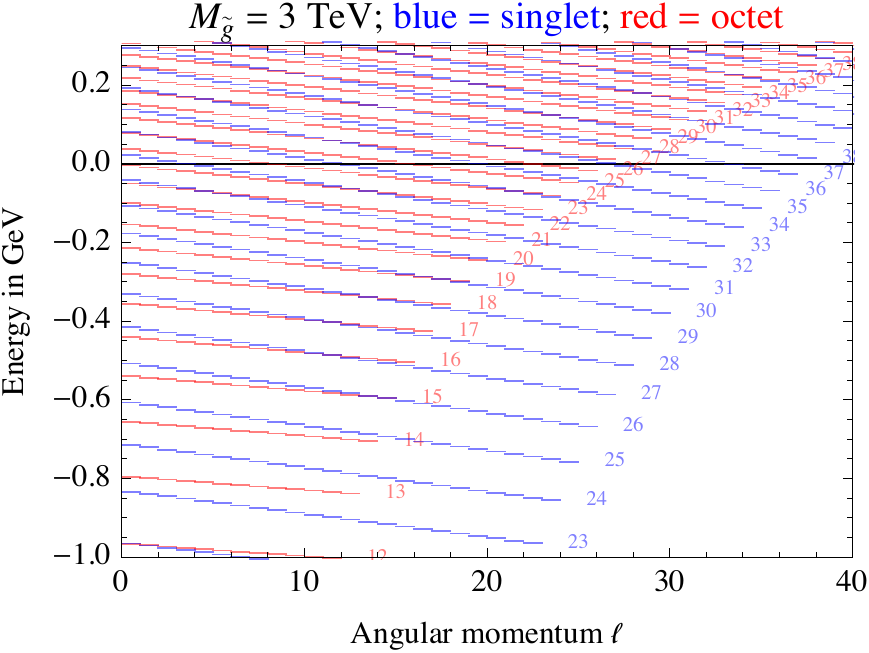}\qquad
\caption{\em \label{fig:Gluinolevels} Quantum energy levels of a $\tilde g\tilde g$ bound state which have energy close to 0.  Values of $n$ are shown.}
\end{center}
\end{figure}

%

Before starting the computations, we summarize generic results for QCD bound states.

\subsubsection*{The bound states}
We compute the energy levels of the $\tilde g \tilde g$ bound states assuming the non-relativistic  QCD potential 
\beq \label{eq:VQCD}
V (r) =\lambda \left\{\begin{array}{lll}\displaystyle
-\frac{\alpha_3(\bar\mu)}{r} \bigg[1 + \frac{\alpha_3}{4\pi} \bigg(\frac{11}{7}+14 (\gamma_E + \ln \bar\mu r)\bigg)\bigg]& r \ll 1/\LQCD &\hbox{\cite{QCDpert}} \\
 \displaystyle
-\frac{\alpha_{\rm 3lattice}}{r} + \sigma r &  r \sim 1/\LQCD&\hbox{\cite{lattice}}
\end{array}\right.\eeq
where $\lambda=(C_R+C_{R'}-C_Q)/2$ for the potential among representation $R$ and $R'$ in the
$Q$ configuration with  $C_1=0$, $C_3=4/3$, $C_8=3$ being the Casimirs.
So $\lambda = 3$ (3/2) for the potential among octets in the singlet (octet)
configuration.
Lattice simulations indicate $\alpha_{\rm 3lattice} \approx 0.3$ and $\sigma \approx (0.4\GeV)^2$.
The one-loop correction to the perturbative term means that the QCD potential is roughly given by the tree level potential with the strong coupling renormalised at  the RGE scale  $\bar\mu\approx 1/r$.
At finite temperature 
$\sigma(T) \approx \sigma(0) \sqrt{1 - \sfrac{T^2}{T_{\rm QCD}^2}}$ with $T_{\rm QCD} \approx 170\MeV$~\cite{lattice}.

The product of two color octets decomposes as
\begin{equation}\label{eq:8x8}
8 \otimes 8= 1_S \oplus 8_A \oplus 8_S \oplus 10_A \oplus \overline{10}_A\oplus 27_S .
\end{equation}
such that there are three attractive channels and the gluino bound states exist in the following configurations
\be
\begin{tabular}{c|ccc|c}
\hbox{Color} & $V$ & \hspace{-3ex}\hbox{i.e.} \hspace{-3ex}& $\lambda$ & \hbox{allowed $\ell$}   \\  \hline
$1_S$ & $-3\alpha_3/r$ && 3 & \hbox{even if $S=0$, odd if $S=1$} \\ 
 $8_A$ & $-\frac 3 2 \alpha_3/r$ && $\sfrac 3 2$& \hbox{even if $S=1$, odd if $S=0$} \\ 
 $8_S$ & $-\frac 3 2 \alpha_3/r$ && $\sfrac 3 2$ &  \hbox{even if $S=0$, odd if $S=1$}  \\ 
\end{tabular}
\label{5boundstates}.
\ee
The energy eigenvalues in a potential $V  = -\alf/r+\sigma_{\rm eff} r$ are~\cite{Hall:1984wk}
\beq \label{eq:enell} E_{ n \ell} \approx \frac{\mu \alf^2}{2} \bigg[-\frac{1}{n^2 t}+12tn \varepsilon x\bigg]\simeq\left\{
\begin{array}{ll}
- \sfrac{\mu \alf^2}{2n^2} & \hbox{Coulomb limit}\\
 \sfrac{3(x\sigma_{\rm eff})^{2/3}}{2\mu^{1/3}}  & \hbox{string limit}
 \end{array}\right.
\eeq
where $\mu \approx M_\Q/2$ is the reduced mass,
$\ell= \{0,1,\ldots\}$ is angular momentum, $n\ge 1 +\ell$,
$x=1.79 ( n- \ell) +\ell -0.42$,
$\varepsilon= \sfrac{\sigma_{\rm eff}}{4\alf^3 \mu^2}$ is a dimension-less  number
and $t$ is the positive solution to
$t=1-4n^3\varepsilon x t^3$.
In the limit where the Coulomb force dominates one has $t\simeq 1$ and $\varepsilon \simeq 0$;
bound states have size $n^2 a_0$ where $a_0 = 1/\mu\alf$ is the Bohr radius.
The  linear force dominates when $n^2 a_0 \gg \sqrt{\alf/\sigma} \sim 1/\LQCD$. 

Fig.\fig{Gluinolevels} shows the energy levels with nearly zero energy for $\Mg=3\TeV$.



\subsubsection*{The breaking rate}
The probabilities $\wp_\ell$ that a given state radiates enough energy before being broken
by a collision can be computed in two different ways.

Based on classical intuition, one can simply compare its energy loss rate
with the breaking rate.
While this simplification holds in the abelian case,
we have to deal with a non-abelian dynamics, where gluon emission changes
singlet to octet states, and vice versa.
This is relevant, as singlet and octet decay rates are significantly different 
(especially for some singlet states which only decay through higher-order effects,
as discussed below).
It's not clear what is the classical limit of this system 
in the limit of large quantum numbers $n,\ell$.  

We then perform a quantum computation, determining the $\wp_\ell$
by simulating transitions among the many different states.
This is feasible up to masses $\Mg\sim 10\TeV$, because it
involves a growing number of states at larger $\Mg$.

We then need the breaking rate of the individual bound states.
Thermal equilibrium  between direct and inverse process (also known as  Milne relation) 
does not allow to infer the breaking rates from the total creation rate, because
the latter is cumulative over all bound states.
We assume that the breaking rate is given by the thermal average of
the pion scattering cross section,
 assumed to be equal to $1/\LQCD^2$,
and perform the thermal average  $\med{\sigma_{\rm break} v_{\rm rel}}$ over the distribution of pions with energies large enough to break the bound states.
The number density of pions with enough energy to break a bound state with binding energy $E_B$ is
 \begin{align}
n_{\pi}^{\rm eq}(E_\pi > E_{B_I})
 \approx \frac{3\,\left(T\, (E_B+m_\pi) \right)^{3/2}}{2 \sqrt{2} \pi^{3/2}} \exp{\left(-\frac{E_B+m_\pi}{T}\right)} \,.
\end{align}
such that $\med{ \Gamma_{\rm break} } \approx \med{ \sigma_{\rm break} v_{\rm rel}} n_{\pi\,}^{\rm eq} (E_\pi > E_{B})$.


\subsection{Color octet states and single gluon emission}\label{8}
We here assume that two colliding $\Q$ can form a $\Q\Q$ system with all 64 possible color
configurations of eq.\eq{8x8},
and with relative weights determined by combinatorics rather than by energetics.
Then the effective annihilation cross section is determined summing over
attractive channels as
\beq\sigma_{\rm ann} \propto
\frac{1}{64} \sigma_{\rm ann}^1 + \frac{1}{8} (\sigma_{\rm ann}^{8_S}
+\sigma_{\rm ann}^{8_A}).
\eeq
We fix the proportionality factor to $\approx 4$ such that the
 total cross section is $\sigma_{\rm QCD} =c/\LQCD^2$, where $c \sim 1$ 
 parameterizes our ignorance of the overall QCD cross section.
%
The annihilation cross section is dominated by $\sigma_{\rm ann}^{8_A}$
because the state $8_A$ radiates much more than $1$ or $8_S$.
Indeed, because of selection rules, single-gluon emission allows the following
decays with $\Delta \ell=\pm1$:
\beq 1\to 8_A, \qquad
8_A\to 1,8_S\qquad
8_S\to 8_A. \eeq
Taking hadronization into account two pions are emitted, such that the binding energy of the final state
$E'_B$ must be larger than $E_B + 2 m_\pi$, otherwise the decay is kinematically blocked.
If the energy gap is somehow bigger than $\LQCD$, inclusive 
decay rates can be  reliably computed treating the gluon as a parton.

Since the 1 state is more attractive than $8_{S,A}$, the above conditions
are  easily satisfied for the $8_A\to 1 $ decay, while $1\to 8_A$ decays 
are kinematically blocked at larger $\ell$ and allowed
at small enough $\ell$ (elliptic enough classical orbit), but
suppressed with respect to the abelian result.

In our numerical results we sum over all possible final states using wave-functions computed
in WKB approximation using the Langer transformation.
We also provide
a simple approximated analytic result obtained assuming Coulombian wave-functions 
(which is valid for deep final states, but not for the QCD-size initial states)\footnote{In the same approximation, the smaller energy radiated into $8_S$ 
is given by a Larmor-like formula,
given that the initial and final state are equally attractive.}
\begin{equation}
\Gamma_{n \ell }(8_A\to 1_S)  \approx \frac 2{n^2}  \alpha_3^5 \mu,\qquad
W_{n\ell}(8_A\to 1_S)= \frac{8 \alpha_3^7 \mu^2 }{n^3 \ell} \,.
\end{equation}

The decay rate must be compared with the thermal breaking rate,
which is given  by pion scatterings such as
$ (\Q\Q) + \pi   \rightarrow (\Q g) + (\Q g)  + \pi$.
Since we considered bound states made of neutral gluinos, 
they are not broken by photon scatterings to leading order.
The result is very simple: the $8_A$ decay rate is so fast that its actual value is irrelevant:
all $8_A$ allowed states have $\wp_\ell=1$ at the relevant temperatures $T \circa{<}\LQCD$.
On the other hand, $8_S$ and $1$ states contribute negligibly.
Then, the annihilation rate is controlled by a much simpler condition:
$8_A$ bound states with binding energy $E_B \sim T$ only exist up to some maximal
$\ell \le \ell_{\rm max8}$, which can be easily computed.
For $M_\Q=3\TeV$ fig.\fig{Gluinolevels} shows that $\ell_{\rm max8}\approx 25$.
For generic $M_\Q \gg T$, $\ell_{\rm max8}$ is well approximated by imposing the vanishing of
$E_{n\ell}$ in eq.\eq{enell}, finding
\beq
\ell_{\rm max8}=(12 \epsilon t^2)^{-1/4} \approx 
\bigg(\frac{3 M_\Q^2 \alpha_3^3}{16\sigma}\bigg)^{1/4}\eeq
having approximated $t\approx 1$ in the last expression.
Using eq.\eq{enell},
the deepest available singlet state has energy gap $\Delta E = \frac94 \sqrt{3\alpha_3 \sigma}\approx 0.9\GeV$ 
(see also fig.\fig{Gluinolevels})
and can only decay via higher order processes.

The effective annihilation cross section  is
\beq\label{eq:suppressionfactor}
 \sigma_{\rm ann} \approx 
\frac{ \sigma_{\rm ann}^{8_A}}{2} \approx \frac12 \sum_{\ell=0}^{\ell_{\rm cr}}
\sigma_\ell
\approx
 \frac12
\frac{2\pi }{M_\Q^2 v_{\rm rel}^2}\ell_{\rm cr}^2,\qquad
\ell_{\rm cr}=\min(\ell_{\rm max},\ell_{\rm max8}).
\eeq
At low (high) temperatures one has $\ell_{\rm cr} \simeq \ell_{\rm max} \propto v_{\rm rel}$
($\ell_{\rm cr} \simeq \ell_{\rm max8}  \propto v_{\rm rel}^0$) such that the thermal average for $\ell\gg1$ is
$\langle \sigma_{\rm ann} v_{\rm rel}\rangle\simeq 2\sigma_{\rm QCD} \sqrt{T/\pi M_\Q}$
($\langle \sigma_{\rm ann} v_{\rm rel}\rangle\simeq \sqrt{3\pi \alpha_3^3/16M_\Q T \sigma}$).
Taking the minimum of these two limits
(which are equal at $T=T_{\rm cr}=\pi \sqrt{3\alpha_3^3/\sigma}/8\sigma_{\rm QCD}$
with $\sigma_{\rm QCD}= \sfrac{c}{\LDC^2} $), 
we obtain an approximation valid at a generic intermediate $T$: 
\beq
\langle \sigma_{\rm ann} v_{\rm rel}\rangle=
\sigma_{\rm QCD}\sqrt{\frac{4T}{\pi M_\Q}}  \left\{
\begin{array}{ll}
0 & \mathrm{for}  \quad T>T_{\rm QCD}\,, \\
 {T_{\rm cr}/T} & \mathrm{for} \quad T_{\rm cr}<T<T_{\rm QCD},\\
1  & \mathrm{for} \quad  T<T_{\rm cr}
\end{array}
\right.
\eeq
\begin{figure}[t]
\begin{center}
\includegraphics[width=.95\textwidth]{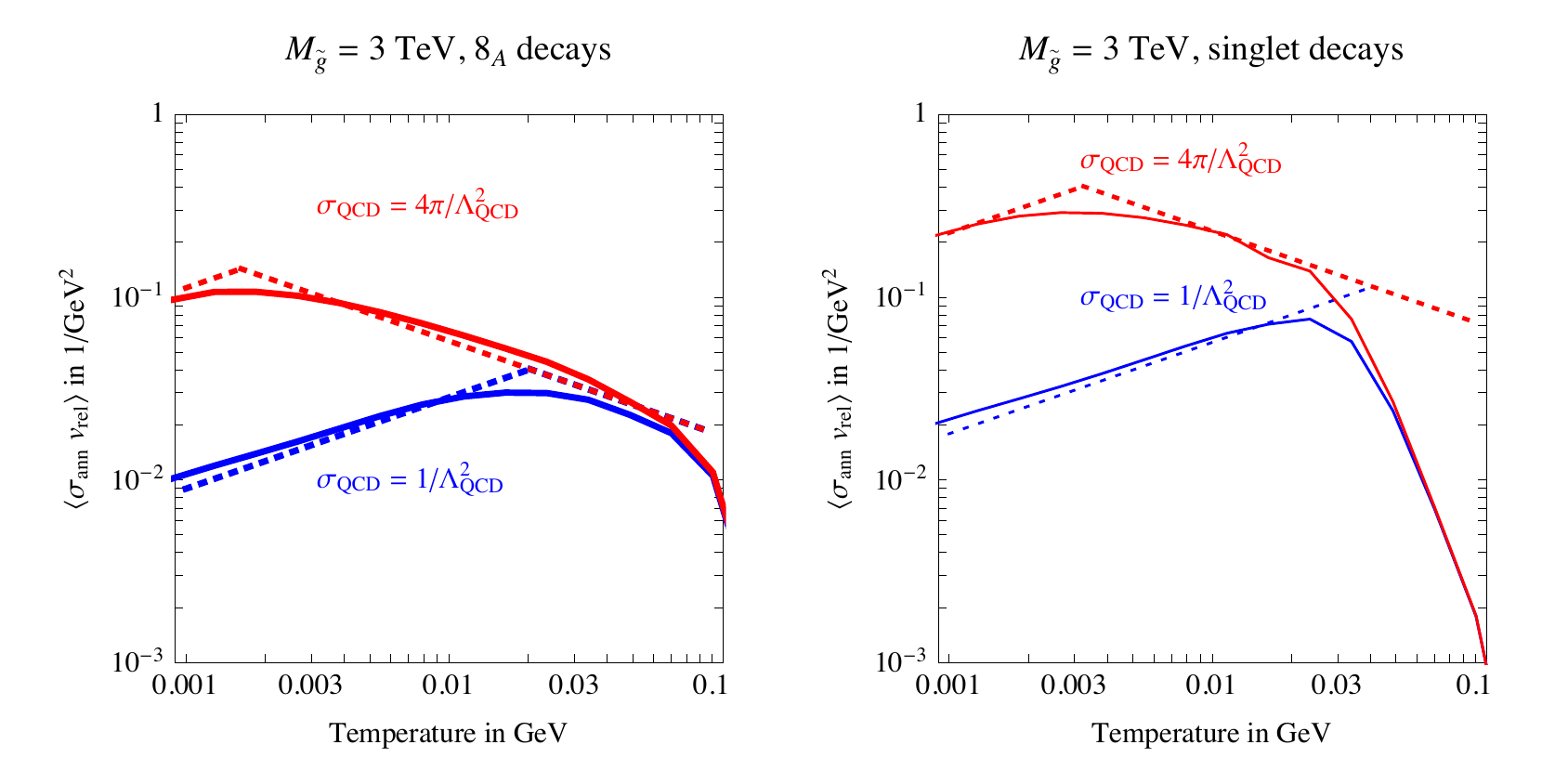}\qquad
\caption{\em \label{fig:SigmaEff} The effective annihilation cross section 
of gluino $\Q\Q$ bound states, assuming that they form color-octet $8_A$ states (left) or
only color-singlet states (right).
The solid curves are the numerical computation the dashed lines are the maximal geometrical cross sections
given by the analytic approximation.}
\end{center}
\end{figure}
The Boltzmann equation of eq.\eq{eff} is approximatively solved by
\begin{equation}\label{eq:Yapprox8}
Y_\Q (\infty)\approx \sqrt{\frac {45}{g_{\rm SM} \pi} }\frac 1 {M_\Q M_{\rm Pl}}\left[ \int_{M_\Q/T_{\rm QCD}}^{\infty} \!\!\! dz\frac{\langle \sigma_{\rm ann} v_{\rm rel} \rangle}{z^2} \right]^{-1}
\approx \frac{9\sqrt{5 M_\Q/g_{\rm SM}} }{4\sigma_{\rm QCD} T_{\rm cr}^{3/2} M_{\rm Pl} 
(3\sqrt{T_{\rm QCD}/T_{\rm cr}}-2)}
\end{equation}
where the $dz$ integral is dominated by $T \sim T_{\rm QCD}$:
for $T_{\rm cr} \ll T_{\rm QCD}$ the abundance simplifies to
\begin{align}\label{eq:Yganal8}
Y_\Q (\infty)\approx \frac{1}{\pi M_{\rm Pl}} \sqrt{\frac{60 M_\Q \sigma}{g_{\rm SM} T_{\rm QCD} \alpha_3^3}}
%
 \approx  0.6~10^{-17} \sqrt{\frac{M_\Q}{3\, \text{TeV}}\frac{170\MeV}{T_{\rm QCD}}} .
\end{align}
The final relic abundance does not have a strong dependence on
$\sigma_{\rm QCD}$, as it is only relevant at relatively low temperatures.  
The DM critical density is exceeded if $M_\Q \circa{>}\PeV$.
Fig.\fig{SigmaEff}a shows the full numerical result for $\med{\sigma_{\rm ann}v_{\rm rel}}$, 
which agrees with the analytic maximal value
(apart from some smoothing at $T \sim T_{\rm cr}$)
up to about $50\MeV$: thereby the numerical abundance is better reproduced
lowering $T_{\rm QCD}$ down to {$50\MeV$} in eq.\eq{Yganal8}.
This is done in the analytic estimate plotted in fig.\fig{Omega}. 

\subsection{Color-singlet states and two gluon emission}\label{1}
Single-gluon emission switches the color of the bound state as $1\leftrightarrow 8$ and its angular momentum
$\ell$ by $\pm 1$:
as a consequence kinematics blocks single-gluon decays of various color-singlet bound states,
roughly all the ones in fig.\fig{Gluinolevels} which don't have nearby octet states.
In particular,  decays of singlet states with maximal $\ell$ are blocked, 
and octet states with maximal $\ell$ can (but need not) decay to singlets with blocked decays.

We thereby take into account 
two-gluon emission, which allows for $1\to 1$ decays with $\Delta\ell = \{0,\pm 2\}$.
The rates of $2g$ transitions are mildly suppressed by
$\mathcal{O}(\alpha_3^3)$ 
compared with the $1g$ decay rates.
If the energy difference $\Delta E$ is much bigger than $\LQCD$, gluon hadronization proceeds
with unit probability and the $2g$ decay widths can be computed using
2nd order non-relativistic perturbation theory~\cite{Bhanot:1979vb}:
\be\label{eq:gamma2}
\begin{split}
\Gamma^{2g}_{n,\ell\to n',\ell'}&\approx  
{\frac{3\alpha_{3}^2}{16\pi}}\int_{0}^{\Delta E}dk\,k^3 \left(\Delta E-k\right)^3\times\\
&\times \sum_{m,m'}\left|\langle \psi_{n,\ell,m}|r_i \left\{\frac{1}{-E_{n',\ell'}+H_8-k}+\frac{1}{-E_{n',\ell'}+H_8-(\Delta E-k)}\right\}r_i |\psi_{n',\ell',m'}\rangle\right|^2
 \end{split}
\ee 
where $r_i = \{x,y,z\}$ is the relative distance between the two $\tilde g$;
$k$ is the momentum of the hadron produced in the hadronization of the two outgoing gluons, 
$\Delta E=E_{n',\ell'} - E_{n,\ell}$ and $H_8$ the free Hamiltonian of the virtual intermediate octet state.
The angular part of the matrix elements, already carried out in eq.~\eqref{eq:gamma2}, imposes the selection rule $|\ell'-\ell |=0,2$.
The two-gluon $1\leftrightarrow 1$ rates are given by an abelian-like expression, unlike the
one-gluon  $1\leftrightarrow 8$ transitions.
The rates for $8\to 8$ two-gluon transitions are given by a similar expression, with $H_8$ replaced by $H_1$.

Hadronization is possible down to the kinematical limit $\Delta E \approx 2 m_\pi$.
However the energy difference between two singlet states with maximal $\ell$,
$|\Delta\ell |= 2$ and  nearby $n$
is $\sim \sigma^{3/4} \alpha_3^{-1/4} \Mg^{-1/2}$, which,
in view of the $\Mg$ suppression, can be smaller than $2m_\pi$.
In such a case the decay can still proceed through off-shell pions, which produce photons and leptons.
We estimate these suppressed decays following section 5.6 of~\cite{1707.05380}.
We neglect multi-gluon emission, which allows bigger jumps in $\ell$.

\bigskip

The $2g$ rates are included 
in numerical computations which assume that QCD-scale color octets exist.
The result was discussed in the previous sub-section,
as $2g$ decays give a relatively minor correction.

We consider the opposite extreme possibility that octet states with QCD-size do not exist,
and that only color singlets exist.
We can again obtain an analytic lower bound on the final $\Q$ abundance by assuming
that all singlet levels fall fast.
Then the cross section $\sigma_{\rm ann} \approx \sigma_{\rm ann}^1$ is only limited by
$\ell_{\rm max1}=\sqrt{2} \ell_{\rm max8}$ such that
\beq
\langle \sigma_{\rm ann} v_{\rm rel}\rangle=
\sigma_{\rm QCD}\sqrt{\frac{16T}{\pi M_\Q}}  \left\{
\begin{array}{ll}
0 & \mathrm{for}  \quad T>T_{\rm QCD}\,, \\
 {T_{\rm cr}/T} & \mathrm{for} \quad T_{\rm cr}<T<T_{\rm QCD},\\
1  & \mathrm{for} \quad  T<T_{\rm cr}
\end{array}
\right.
\eeq
where now $T_{\rm cr}=\pi \sqrt{3\alpha_3^3/\sigma}/4\sigma_{\rm QCD} $.
The resulting relic gluino abundance is 2 times lower than in eq.\eq{Yapprox8},
and with the new value of $T_{\rm cr}$.
Fig.\fig{SigmaEff}b shows that this limit only holds at $T \circa{<}20\MeV$,
such that the analytic expression reproduces the numerical value for $Y_{\Q}$
by reducing $T_{\rm QCD}$ down to $\sim 20\MeV$.

 \begin{figure}[t]
\begin{center}
\includegraphics[width=.45\textwidth]{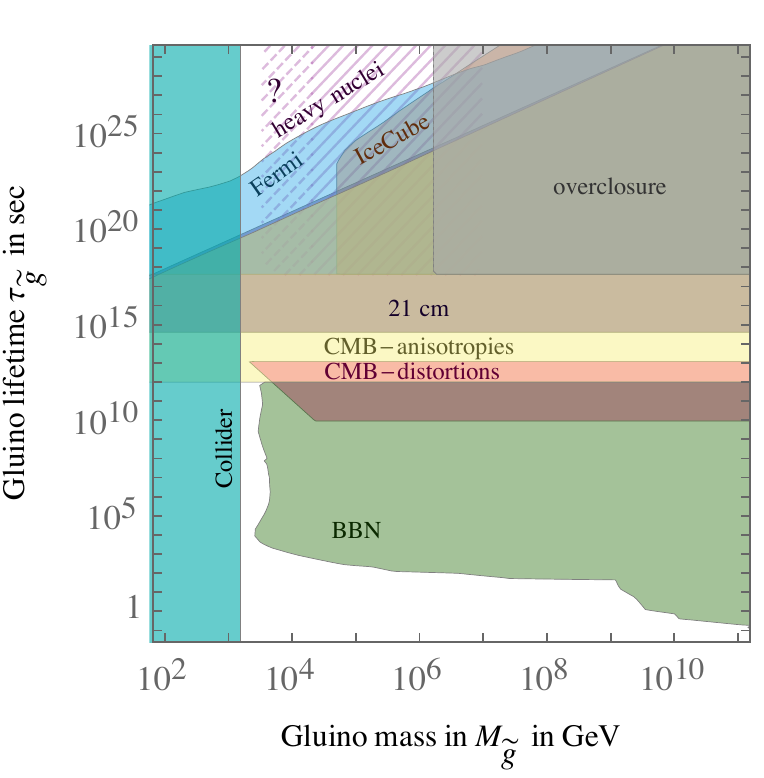}\quad
\includegraphics[width=.45\textwidth]{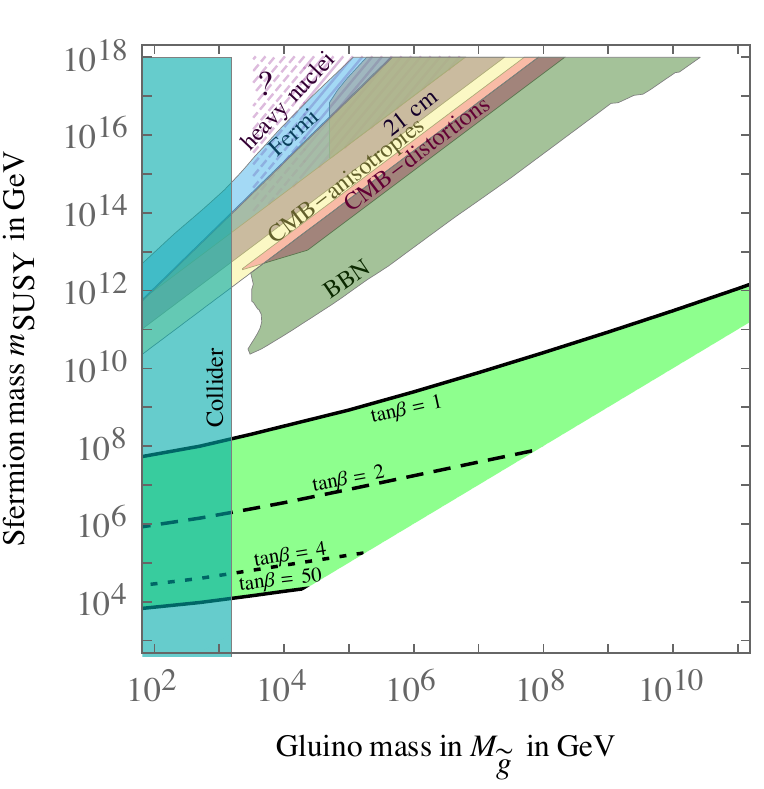}
\caption{\em\label{fig:constraints}Cosmological constraints on long-lived gluinos.
{\bf Left}: As a function of the gluino lifetime.
{\bf Right}: As a function of the sfermion mass scale $m_{\rm SUSY}$, which in Split SuperSymmetry
determines the gluino lifetime. }
\end{center}
\end{figure}

\section{Phenomenology}\label{pheno}

\subsection{Cosmological bounds and signatures}
Bounds on quasi-stable relics depend on their lifetime $\tau_\Q$;
on their mass $M_\Q$;
on their relic abundance, that for gluinos we computed in terms of $M_\Q$, and on their decay modes.
As mentioned above, we assume that gluinos decay to neutralinos (assumed to be
the Lightest Super-symmetric Particle, LSP)
plus either a gluon or a quark and an antiquark.
Here we assume that half of gluino energy is carried away by the LSP; if the LSP is not much lighter than the gluino, even less energy goes into SM states and one would obtain weaker bounds.

Our final result is plotted in fig.\fig{constraints}, using the thick red dashed line of fig.\fig{Omega}: 
even using updated experimental bounds (discussed below),
our bounds on a (quasi)stable gluino
are significantly weaker than those derived in~\cite{Mina}.
The reason is that our relic density takes into account non-perturbative gluino annihilations, 
and is much smaller than the  `perturbative' gluino relic density assumed in~\cite{Mina},
see fig.~\ref{fig:Omega}.
In particular, we find that a (quasi)stable gluino just above present collider bounds 
is still allowed provided that its lifetime is smaller than  about $10^{12}\s$
or larger than about $10^{22}\s$.

In the rest of this section we
summarize the various bounds on decaying relics plotted in fig.\fig{constraints}, 
moving from smaller to larger lifetimes.

\subsubsection*{Big Bang Nucleosynthesis}

A gluino that decays during BBN can disturb the successful BBN predictions of light element abundances,
which get affected in different ways, depending on the gluino lifetime (for more details see \cite{Kawasaki:2004qu,Kawasaki:2017bqm}):\footnote{In addition, gluinos could also disturb the BBN predictions if they participate themselves in the nuclear reactions occurring during BBN~\cite{Pospelov:2006sc,Kusakabe:2009jt}.
This would be the case if the gluino $R$-hadrons bind into nuclei which are relevant during BBN.
Since we do not know whether this is the case or not, we ignore such effects here.}
\begin{itemize}
\item
For $0.1 \s\lesssim \tau_{\tilde g} \lesssim 10^2 \s$ the mesons and nucleons produced by  gluino decays quickly reach kinetic equilibrium with the thermal bath of background photons and $e^\pm$ and thus do not have enough energy to destroy light nuclei.
However, the extra pions, kaons and nucleons present in the thermal bath increase the $p \leftrightarrow n$ conversion rate, thus increasing the $n/p$ ratio and as a consequence the primordial $^4$He mass fraction $Y_p$.

\item
For $\tau_{\tilde g} \gtrsim 10^2\s$ the gluino decay products do not thermalize before interacting with nuclei, due to the lower temperature of the plasma at these times. The still energetic nucleons (the mesons decay before they can interact) can thus hadrodissociate $^4$He which in turn also increases the D abundance (e.g. via $p+ {}^4\hbox{He}\to \hbox{D}+{}^3\hbox{He}$).

\item
For $\tau_{\tilde g} \gtrsim 10^7\s$ photodissociation of $^4$He, which induces increased $^3$He and D abundances, becomes relevant.
Photodissociation is not relevant at earlier times because the $\gamma$-spectrum is cut off at the threshold energy $E_{\rm th}^\gamma \approx m_e^2/(22\, T)$~\cite{Kawasaki:1994sc} for $e^+ e^-$ pair production from energetic $\gamma$'s with thermal $\gamma$'s, so that photons are not energetic enough to break up nuclei.

\end{itemize}
The resulting constraints have been computed in~\cite{Kawasaki:2004qu} and updated and improved in \cite{Kawasaki:2017bqm}.
The constraints are given in the $(\tau_{X}, \xi_X)$ plane for different main decay modes of $X$, where $X$ is the unstable relic (the gluino in our case) and $\xi_X = E_{\rm vis} Y_X$ is its destructive power.
Since we assume that half of gluinos' energy is carried away by the LSP we have $E_{\rm vis} \approx M_\Q/2$.
The bounds for the various hadronic decay modes are similar since in all cases they induce hadronic showers, and our bounds are based on the plot for the $t \bar t$ mode.

The effects from photodissociation depend only on the total injected energy, so that for $\tau_{\tilde g} \gtrsim 10^7\s$ the bounds do not explicitly depend on $M_\Q$ to a good approximation.
At earlier times, the effects depend on the number of hadrons produced in the hadronization process, which scales with a power of $M_\Q$.
Thus we fit the bounds, given in \cite{Kawasaki:2017bqm} for $M_X=1 \TeV, 10 \TeV, 10^2 \TeV, 10^3\TeV$, to a power-law function of $M_\Q $.

The left-handed panel of fig.~\ref{fig:constraints} shows the resulting bounds in green.
In the right-hand panel we show the same bounds with the gluino lifetime computed as
function of the SUSY breaking scale $m_{\rm SUSY}$.

\subsubsection*{Distortion of the CMB blackbody spectrum}
Gluinos with lifetimes between $\sim 10^7 \s$ and $\sim 10^{13}\s$ (the latter corresponds to recombination) can lead to deviations of the CMB spectrum from a blackbody form.
When the Universe is $10^7 \s$ old, photon number changing processes such as double Compton scattering are not efficient any more, so that photons injected into the plasma can induce a chemical potential $\mu \simeq 1.41 \ \delta \epsilon/\epsilon$~\cite{Wright:1993re} 
in the Bose-Einstein distribution of the CMB radiation, where~\cite{Hu:1993gc}
\begin{align}
\frac{\delta \epsilon}{\epsilon} \simeq 4 \times 10^{-3} \sqrt{\frac{\tau_{\tilde g}}{10^6 \s}} \frac{M_\Q  Y_{\tilde g} B_\gamma}{10^{-9} \GeV} \exp\left[-\left(\frac{6.1 \times 10^6 \s }{ \tau_{\tilde g}}\right)^{5/4}\right] \,.
\end{align}
After $\sim 4 \times 10^{11} \Omega_b h^2 \s$~\cite{Hu:1993gc}, elastic Compton scatterings do not maintain thermal equilibrium anymore.
An injection of photons `Comptonizes' the spectrum, i.e.\ 
it leads to a mixture of blackbody spectra of different temperatures.
This is described by the Compton $y$-parameter, given by $y= \delta \epsilon/4\epsilon$~\cite{Wright:1993re}.

The 95\% CL limits from the FIRAS instrument on the COBE satellite are $|\mu|< 9 \times 10^{-5}$ and $|y|<1.5 \times 10^{-5}$~\cite{Mather:1993ij,Fixsen:1996nj}. 
The resulting constraints on the gluino lifetime are shown in pink in fig.~\ref{fig:constraints}. 
Here we assumed that $\sim$~45\% (see e.g.~\cite{Cirelli:2010xx}) of the energy that is not carried away by the LSP goes into photons.
The resulting bounds are less constraining than the BBN bounds.
However future bounds from PIXIE~\cite{Kogut:2011xw} will be stronger by 2 to 3 orders of magnitude.

\subsubsection*{CMB anisotropies}

The electromagnetic energy ejected into the gas at or after recombination by decaying relics modifies the fraction of free electrons and heats the intergalactic medium.
This leads to modifications of the CMB angular power spectrum,  measured by {\sc Planck}. 
The maximally allowed density of a long-lived relic as a function of its lifetime has been computed assuming decay products with fixed energies in the range from 10 keV up to 10 TeV~\cite{Slatyer:2016qyl} respectively 1 TeV~\cite{Poulin:2016anj}.
The $e^+,e^-,\gamma$ from hadronic decays do not have fixed energies, and moreover we do not know the energy spectrum of the decay products of relics with a mass significantly larger than 10 TeV.
For very large gluino masses the bounds we show are therefore only indicative.
We consider the middle of the band in~\cite{Poulin:2016anj} and obtain bounds by assuming that half of gluinos energy goes into SM states and that 60\% (see e.g.~\cite{Cirelli:2010xx}) of the latter goes into $e^+,e^-,\gamma$.
In fig.~\ref{fig:constraints} we show the resulting constraints for a gluino with a lifetime $\gtrsim 10^{12} \s$ in yellow.

\subsubsection*{21-cm line}
If confirmed, the observation of an absorption feature in the low energy tail of the CMB spectrum \cite{Bowman:2018yin} allows us to put an upper bound on the temperature of the intergalactic medium (IGM) at redshift $z\approx17$. 
Decays of relic particles during the dark ages are  constrained, mainly because
they inject energy in the IGM  heating it, erasing the  absorption feature.
Bounds on decaying DM particles, with masses up to $10\TeV$, have been computed in \cite{1803.09739,1803.09390,1803.11169}. 
We rescale these bounds to a generic abundance, still assuming
that half of gluino energy goes into SM states and that 60\% (see e.g.~\cite{Cirelli:2010xx}) 
of the latter goes into $e^+,e^-,\gamma$.
The result is shown in fig.~\ref{fig:constraints}. 
Similarly to the case of the CMB bounds in the previous section, the 21 cm bounds for very large gluino masses are only indicative and subject to significant uncertainty. 

\subsubsection*{Constraints from gamma-ray telescopes and neutrino detectors}
Decaying gluinos with larger lifetimes are constrained by the measurement of
cosmic ray spectra, in particular of photons of neutrinos.
We adopt the results of~\cite{Cohen:2016uyg} who computed 
limits on the lifetime of DM decaying to $b \bar b$, 
from data from the {\sc Fermi} gamma ray telescope and 
the neutrino detector {\sc IceCube}, up to a DM mass of $10^{12}$ GeV.
We rescale the bounds of~\cite{Cohen:2016uyg} taking into account that the density of our relics
differs from the DM density.
Ref.~\cite{Cohen:2016uyg} derives bounds assuming a relic that decays to $b \bar b$. We assume that 50\% of the gluino's energy goes to the LSP and the rest goes into hadronic decay channels, which lead to similar spectra as $b \bar b$.
Fig.~\ref{fig:constraints} shows the resulting constraints on a long-lived gluino from {\sc Fermi} (in blue) and
{\sc IceCube} (in orange).
The {\sc IceCube} limits exceed the bounds from {\sc Fermi} data for  $M_\Q \gtrsim 10^7$ GeV.

\subsubsection*{Searches for super-massive nuclei}
Coming finally to stable gluinos, lattice simulations indicate that they would 
form neutral $\tilde g g$ hadrons~\cite{Foster:1998wu},
as well as a minor component of baryonic states
such as $\Q uud$ 
(according to~\cite{1011.2964} the lightest gluino baryon could be $\Q uds$).
They behave as strongly interacting Dark Matter.
This is allowed by direct detection experiments performed in the upper atmosphere
and by searches for super-massive nuclei in the Earth and in meteorites 
if their relic abundance is a few orders of magnitude
smaller than the cosmological DM abundance,
although the precise bound is subject to considerable uncertainties
(see the discussion in~\cite{1801.01135}).
In fig.\fig{constraints} we indicate the tentative constraints that arise from the search for supermassive nuclei in meteorites by Rutherford backscattering of $^{238}$U, $N_{\textrm{SIMP}}/N_n\vert_{\textrm{meteorites}} \lesssim 2 \times 10^{-12}$~\cite{Polikanov:1990sf}, assuming a heavy nuclei capture cross section of $\sigma_{\textrm{capture}}=10^{-2}/\Lambda_{\textrm{QCD}}^2$.
Presumably, there is still an open window, from
TeV masses above the LHC~\cite{hep-ph/0611322} up to about 10 TeV.

\subsubsection*{Higgs mass}
In the right panel of fig.\fig{constraints} we considered Split SuperSymmetry,
such that the gluino lifetime is computed as function of the sfermion mass $m_{\rm SUSY}$, see eq.\eq{taug}.
This scale is further constrained within the split MSSM by the 
observed Higgs mass, which is reproduced within the green region (for different values of $\tan\beta$) in the $(M_3, m_{\rm SUSY})$ plane.  We computed $M_h$ as in~\cite{1407.4081},
assuming that gauginos and Higgsinos 
are degenerate at the gluino mass $M_3$ and that all scalars are degenerate at $m_{\rm SUSY}$.   Allowing the masses to vary and taking into account uncertainties on $M_t$ and $\alpha_3$ slightly expands the region.
Within the Higgs-allowed region the gluino decays promptly on cosmological time-scales, evading all cosmological bounds.

No prediction for the Higgs mass arises in extensions of the MSSM.
However, roughly the same region is obtained imposing the meta-stability bound on Higgs vacuum decay,
which implies that the Higgs quartic $\lambda_H$ cannot be too negative, $\lambda_H \circa{>}-0.05$.
A substantially larger $m_{\rm SUSY}$, such that the gluino is long-lived, is obtained assuming that
Higgsinos are heavy (possibly with masses of order $m_{\rm SUSY}$: in such a case the RGE for
the Higgs quartic are those of the SM (with slightly different values of $g_{2,3}$ due to the light gluino and wino), and the Higgs quartic can remain positive up to $m_{\rm SUSY}\sim M_{\rm Pl}$ within the uncertainty range for the top quark mass.

\subsection{Collider signals}
Next, we discuss some aspects of the phenomenology of long-lived gluinos at hadron colliders,
in particular LHC.
Long-lived gluinos can be pair produced and after hadronization form long-lived
hybrid states with SM quarks and gluons, known as `$R$-hadrons'. 
We conservatively assume that the signal at the LHC is just energy deposit in the calorimeter,
rather than charged particles in the tracker.
It is difficult to trigger on these event and so an initial state jet is required. The LHC places the limit
$M_\Q > 1.55 \text{ TeV}$ on a Majorana gluino~\cite{ATLAS:2018yey}. 

The other possibility is the production of a $\Q\Q$ bound state.
Assuming that states with $\ell=0$ dominate the rates,
they are color $8_A$ with spin $S=1$ and
color singlets or $8_S$ with $S=0$ (see eq.~\ref{5boundstates}).
The production cross sections are given by gluon and quark fusion respectively 
\begin{align}
& \sigma_0 = \sum_{n=1}^\infty \frac{\mathcal{L}_{gg} }{2 M_\Q \,s\, n^3} \left( \Gamma_{gg}^1 + 8 \Gamma_{gg}^{8_S} \right) = \frac{ \zeta(3) \, \mathcal{L}_{gg} }{2 M_\Q \,s} \left( \Gamma_{gg}^1 + 8 \Gamma_{gg}^{8_S} \right)\,,\\
& \sigma_1 = 2 \sum_{n=1}^\infty \frac{ \mathcal{L}_{uu}  \Gamma_{uu}^{8_A} +  \mathcal{L}_{dd}  \Gamma_{dd}^{8_A}}{ M_\Q \,s\, n^3}  =   \frac{ 2\, \zeta(3)  }{M_\Q \,s} \left( \mathcal{L}_{uu}  \Gamma_{uu}^{8_A} +  \mathcal{L}_{dd}  \Gamma_{dd}^{8_A} \right)\,,
\end{align}
where ${\cal L}_{ij}$ is the luminosity of partons $ij$.
The decay rates are given by \cite{1801.01135}
\begin{align}
\frac{\Gamma_{gg}^1}{M_\Q} = \frac{9 \alpha^5_3 \lambda_1^3}{2 \,F}\,, 
\qquad \frac{\Gamma_{gg}^{8_S}}{M_\Q} = \frac{9 \alpha^5_3 \lambda_8^3}{8 \,F}, \qquad \frac{\Gamma_{qq}^{8_A}}{M_\Q} = \frac{3 \alpha^5_3 \lambda_8^3}{2 \,F} \,  ,
\end{align}
with $F =2$ for the Majorana gluino and $F=1$ for a Dirac particle, and with the channel strength $\lambda_1 =3 $ and $\lambda_8 =3/2$.  

\begin{figure}
\begin{center}
\includegraphics[width=.5\textwidth]{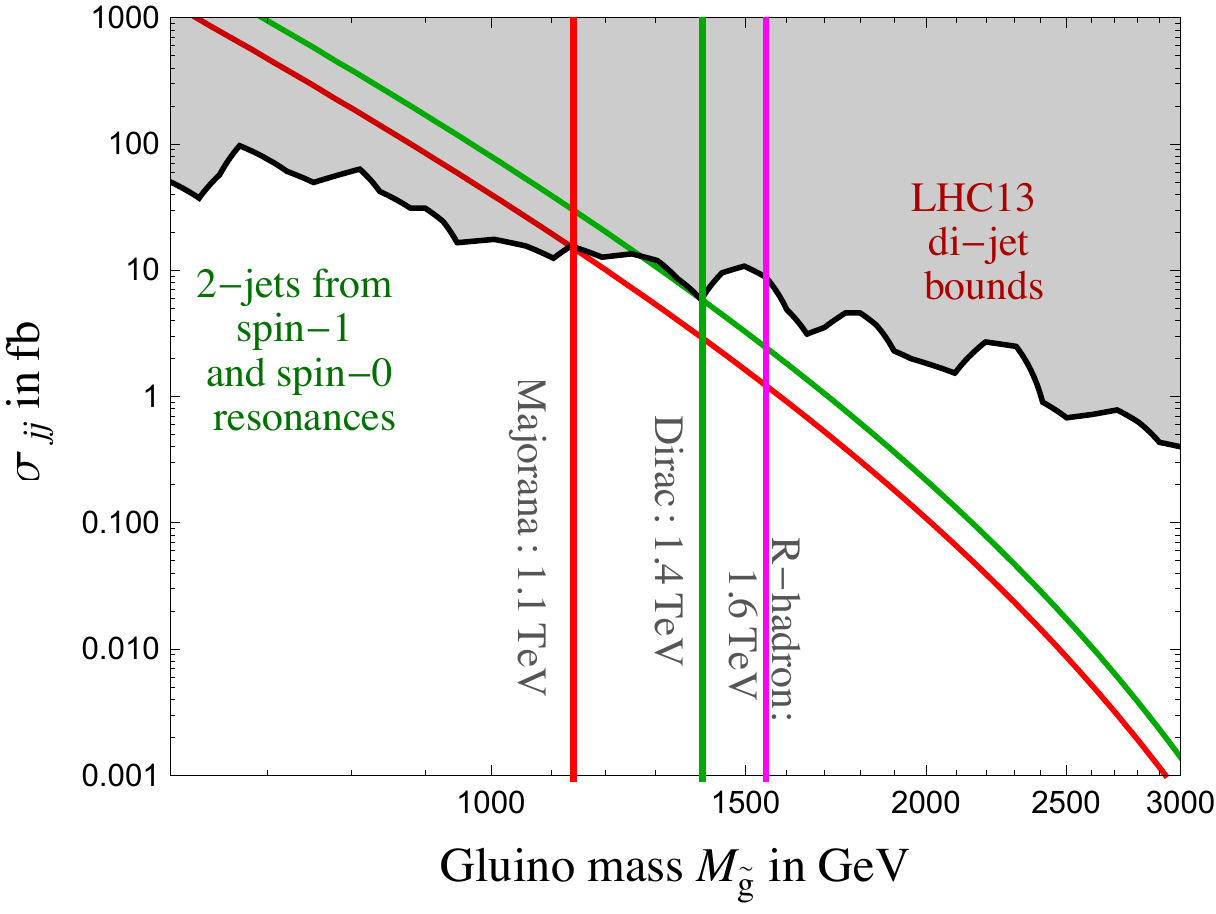}\qquad
\caption{\em\label{fig:LHC}
The black curve is the di-jet upper bound on the cross section for
production of spin-1 and spin-0 bound states from LHC data at $13\TeV$;
the red (green) curve is the theoretical prediction assuming a
Majorana (Dirac) gluino. 
From this we derive the experimental bounds (vertical lines).
The thin vertical line shows the bound 
from $R$-hadron searches.}
\end{center}
\end{figure}

Since the resonances annihilate to two gluons or two quarks, we assume a $100\%$
branching ratio  to two jets and apply the LHC di-jet bounds \cite{di-jet} to the sum of the cross sections. 
In fig. \ref{fig:LHC} we compare the bounds on the resonances to, slightly stronger, the $R$-hadron bound.

Concerning future colliders, the expected 
reach of a 100 TeV hadron collider with $1000 \,\text{fb}^{-1}$ is 7 (9)TeV for a Majorana (Dirac) gluino,
having used~\cite{1402.0492} to perform an approximate rescaling.
The $R$-hadron search would then reach $10$ TeV and $14.5$ TeV respectively.  
Thus a 100 TeV collider would reach the benchmark mass of a thermally produced Dirac gluino, which recently was found to be a dark matter candidate \cite{1801.01135}. 


 \begin{figure}[t]
\begin{center}
\includegraphics[width=.81\textwidth]{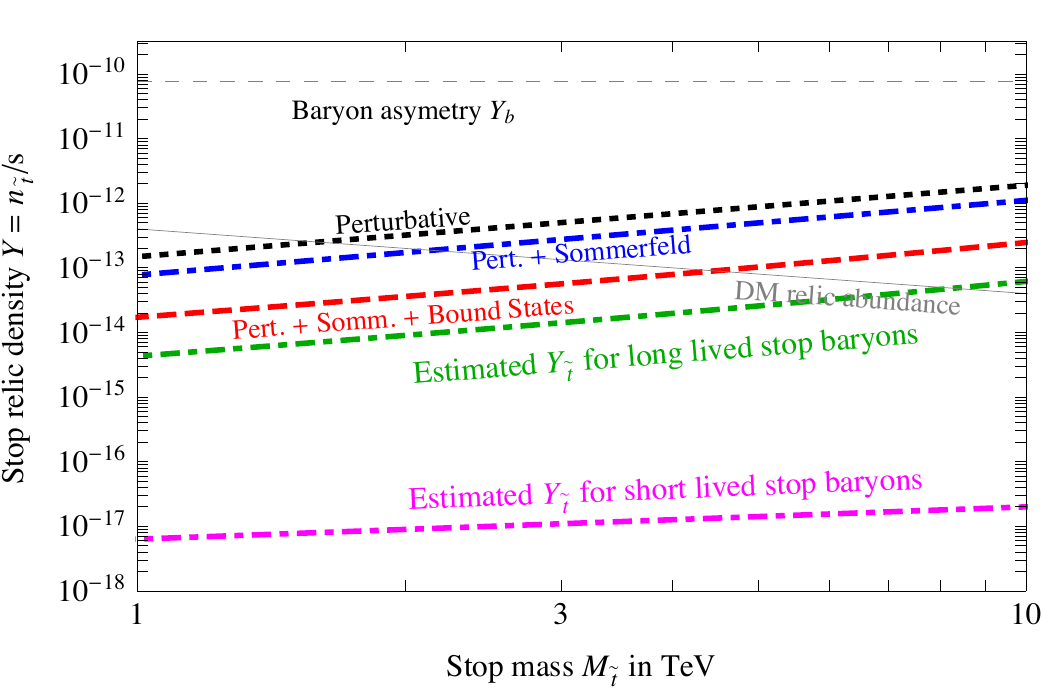}\qquad
\caption{\em\label{fig:stop}
Stop relic abundances. The $\tilde t\tilde t\tilde t$ baryons could be
relatively long lived and have an abundance not suppressed by QCD confinement effects
}
\end{center}
\end{figure}

\subsection{Implications for Dark Matter co-annihilations}
\begin{figure}[t]
\begin{center}
\includegraphics[width=.47\textwidth]{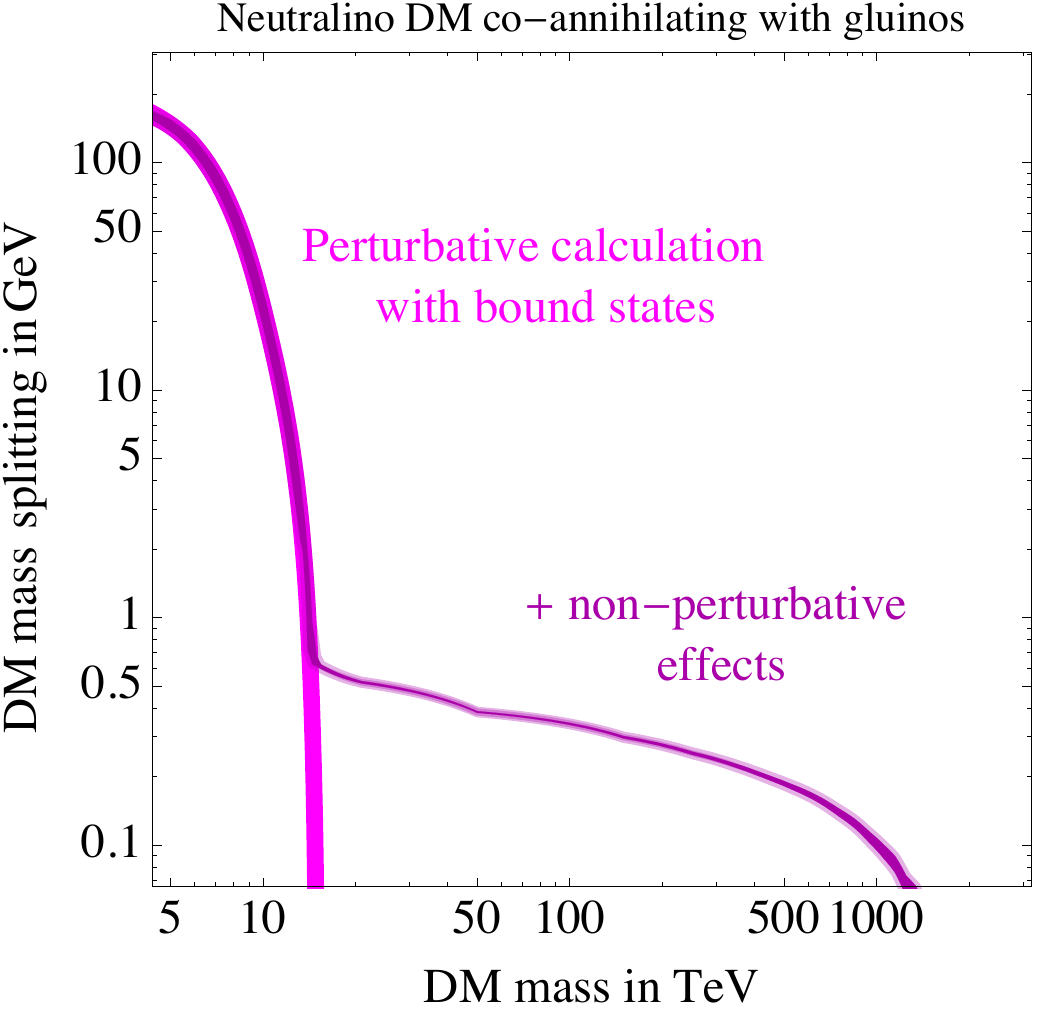}\qquad
\includegraphics[width=.47\textwidth]{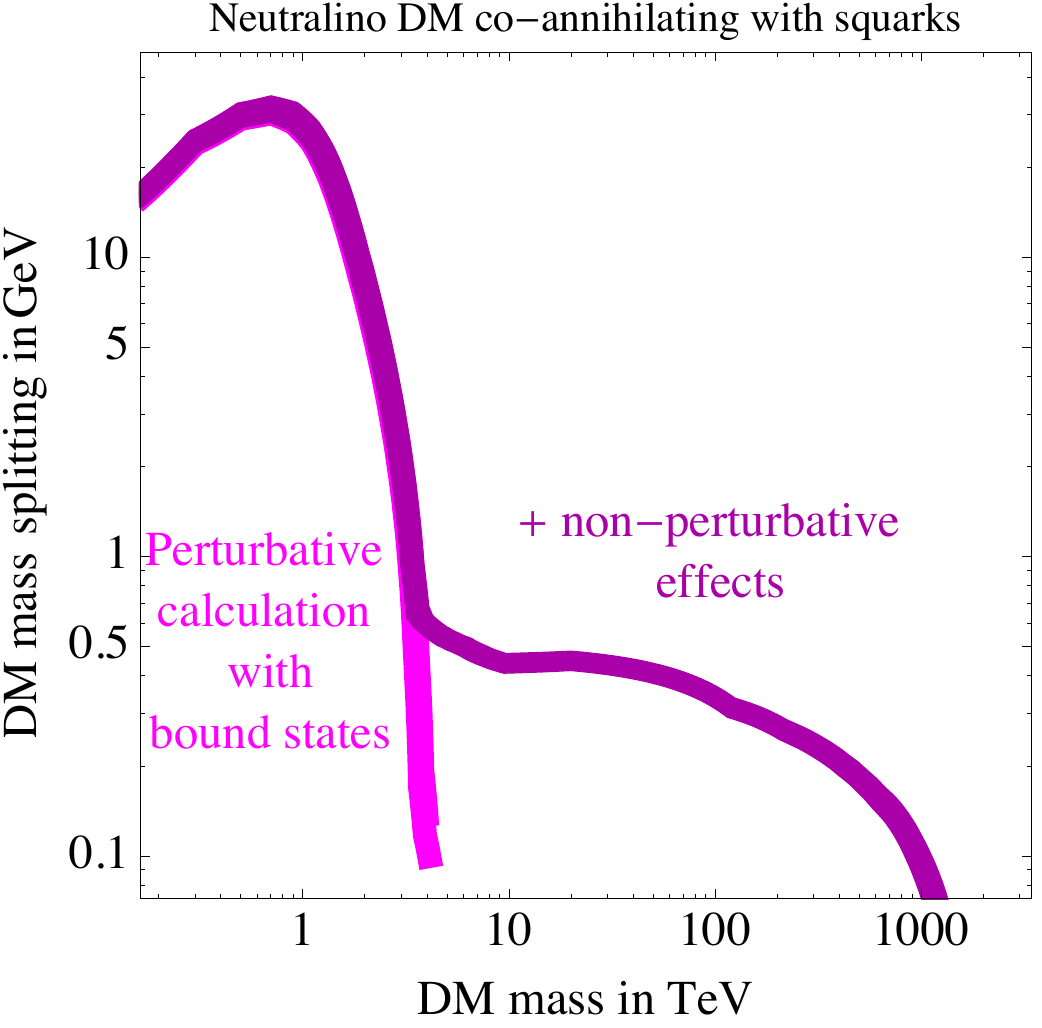}
\caption{\em \label{fig:GluinoCAN} 
Non-perturbative QCD annihilations that take place at $T \circa{<}\LQCD$ 
significantly increase the DM neutralino mass such that the observed DM abundance is reproduced trough
co-annihilations with gluinos (left) or stops (right), 
if their mass difference with neutralinos is smaller than a few $\GeV$. 
In the case of stops (right panel), the big effect is only estimated and only present 
if stop baryons decay to SM particles before decaying to neutralinos;
otherwise confinement only gives a ${\cal O}(1)$ effect.}
\end{center}
\end{figure}

The thermal relic abundance of a particle is affected by co-annihilations with particles of similar mass.
One example is co-annihilations of neutralino DM with heavier colored particles, for example gluinos.
Co-annihilations can be enhanced by Sommerfeld corrections~\cite{1402.6287} and bound-state formation ~\cite{Ellis:2015vaa,1702.01141}.
We point out that a much bigger effect is produced by non-perturbative QCD effects after the QCD phase transition, if the mass splitting $\Delta M$ between the co-annihilating species is comparable or smaller than
$\LQCD$.
Such a near-degeneracy is unnatural.
This is shown in fig.~\ref{fig:GluinoCAN}a in the neutralino/gluino co-annihilation case,
assuming that squarks mediate fast neutralino/gluino rates.
We see that the neutralino mass which reproduces the observed DM density gets much higher
at  $\Delta M \circa{<} \GeV$.
In the limit $\Delta M \ll \GeV$ the relic abundance is dominantly set by the new QCD annihilations.
As a result,  the neutralino mass can reach up to a PeV, 
heavier than the maximal relic DM mass allowed if DM annihilations are dominated
by partial waves with low $\ell$~\cite{Griest:1989wd}.

\subsection{Quasi-stable squark}\label{triplet}
In the previous sections we considered a Majorana gluino.
A real scalar in the octet of $\SU(3)_c$ would behave similarly to the Majorana gluino.
On the other hand, a (quasi)stable
particle in the fundamental 3 of color $\SU(3)_c$ can behave in a qualitatively different way.
Since the 3 is a complex representation, the particle must be a complex scalar or a Dirac fermion,
which can carry a conserved charge.

For definiteness, we consider the possibility of a (quasi)stable squark,
and more specifically a stop $\tilde t$, as RGE effects tend to make $\tilde t$
lighter than other squarks.
A stable stop arises if $\tilde t$ is the lightest SUSY particle (LSP) and $R$-parity is conserved.
A quasi-stable stop arises if $R$-parity is almost conserved,
or if the stop decays slowly into the LSP:
this can happen e.g.\ when the LSP is a gravitino.
Collider bounds on stops~\cite{1212.6847} tend to ignore the possibility 
that the lighter stop $\tilde t$ is the (quasi)stable LSP,
because it is perceived to be already excluded by cosmology.

In cosmology, perturbative QCD $\tilde t\tilde t^*\to gg$
annihilations dominate over $\tilde t \tilde t \to tt$ annihilations 
and leave a roughly equal amount of relic $\tilde t$ and $\tilde t^*$.
Perturbative QCD annihilations are enhanced by Sommerfeld and bound-state effects,
computed in \cite{1702.01141}. The relic $\tilde t $ abundance after perturbative
annihilations is plotted in fig.\fig{stop} 
and approximated by
\beq \frac{n_{\tilde t}}{s} \approx  \frac{M_{\tilde t}}{M_{\rm Pl} \alpha_3^2}.\eeq
For $M_{\tilde t} < \PeV$ this is smaller than the baryon asymmetry $n_b/s \sim 10^{-10}$, that we neglect given that its effect is model dependent.
Indeed, we do not know how the baryon asymmetry is generated: it might be generated at the weak
scale such that it would not affect heavier stops.
Even if a baryon asymmetry is present at stop decoupling, 
$\tilde t \bar t \leftrightarrow \tilde t^* t$ scatterings 
could easily concentrate the baryon asymmetry to lighter 
baryons fast enough that the asymmetry is irrelevant for stops.
If instead the baryon asymmetry enhances the relic stop abundance, bounds would become stronger.

\medskip

After the QCD phase transition, stops form hadrons.
In view of the large QCD cross sections, the stop hadrons with dominant abundance
are deeply-bounded states which contain  stops only.
They are $\tilde t\tilde t^*$ and the charged baryons $\tilde t\tilde t \tilde t$.
Both fall to the ground state and decay through annihilations of the constituents.
In particular, a bound state containing two or more stops
decays, in its ground state, 
with a life-time $\Gamma_{\tilde t \bar t}  \sim \alpha_3^3 M_{\tilde t}^3 \sigma_{\tilde t \bar t} v_{\rm rel}$ where 
the cross section for $\tilde t\tilde t \to tt$ can be roughly estimated as
$\sigma_{\tilde t \bar t} v_{\rm rel}\sim \sum_{i=\{1,2,3\}} \alpha^2_i/ M_{i}^2$,
ignoring possible extra velocity suppressions.
Then, $\Gamma_{\tilde t \bar t} $ is cosmologically fast unless
gauginos (with masses $M_i$) are heavier than $\sim 10^{10}\GeV$.


We expect a roughly equal number of 
$\tilde t\tilde t^*$ annihilations for each produced
$\tilde t\tilde t \tilde t$
given that QCD group algebra implies that both $\tilde t\tilde t^*$ and $\tilde t\tilde t$ feel an attractive Coulombian
QCD force, such that they can form deep, unbreakable, Coulombian bound states.
Assuming that a $\tilde t$ binds with probability $\wp$ to a $\tilde t$ 
and with probability $1-\wp$ to a $\tilde t^*$ 
and  thereby that a deep $\tilde t\tilde t$ binds with probability $1-\wp$ to  $\tilde t$
and with probability $\wp$ to a $\tilde t^*$,
the average number of $\tilde t \tilde t^*$ per produced baryon is
\beq 
\frac{\langle N_{\tilde t \tilde t^*} \rangle}{\langle N_{\tilde t\tilde t\tilde t}+N_{\tilde t^*\tilde t^*\tilde t^*} \rangle}  =   \frac{1/\wp+1/(1-\wp)-1}{r+1/r-1}.
\eeq
This equals $3$ assuming no baryon asymmetry $r \equiv N_{\tilde t}/N_{\tilde t}^*$ 
and $\wp=1/2$, namely neglecting that $\tilde t\tilde t^*$ is more attractive than $\tilde t\tilde t$.
Extra hadrons and mesons that contain quarks have a much smaller abundance,
that is not relevant here.
If the charge 2 states $\tilde t\tilde t \tilde t$ decay fast on cosmological scales,
final abundances and bounds are similar to the gluino case.
If (quasi)stable, they are instead subject to strong  cosmological constraints.
In particular during BBN $\tilde t^* \tilde t^* \tilde t^* $ can bind to $^4{\rm He}$ 
reducing its charge and thereby the Coulomb suppression of nuclear reactions,
opening up a new channel for $^6$Li production, 
\begin{equation}
	(\tilde t^* \tilde t^* \tilde t^* \ \mathrm{^4He}) + D \to \mathrm{^6Li} + \tilde t^* \tilde t^* \tilde t^* \,,
\end{equation}
which can  strongly alter Lithium abundances (see~\cite{1011.1054} for a brief review).
Charge $-1$ states with  lifetime $\gtrsim 10^5$ are subject to the BBN bound
$Y \lesssim 2.5 \times 10^{-17}$~\cite{Pospelov:2006sc}.
A study of analogous constraints on relics with charge $-2$  is beyond the scope of this paper.

\medskip
\medskip

Next, we study the scenario where a quasi-stable stop co-annihilates
with a slightly lighter DM neutralino.
Post-confinement effects are relevant if $\Delta M \circa{<}\GeV$.
Roughly half of the stops form $\tilde t \tilde t^*$ mesons,
and the others form $\tilde t\tilde t \tilde t$ baryons.
The impact on the DM abundance is very different, depending on 
which process dominates $\tilde t\tilde t \tilde t$ decays.
If it is dominated by stop annihilations into SM particles, 
post-confinement effects strongly suppress the DM abundance,
similarly to the  gluino/neutralino co-annihilation  scenario.
A much smaller order one effect is obtained if instead
stops decay to DM neutralinos and SM particles with rate $\Gamma_{\tilde t}\circa{>} \Gamma_{\tilde t \tilde t}$.
The region where the DM abundance is reproduced is estimated 
in fig.\fig{GluinoCAN}b in the two extreme possibilities, having assumed
$\sigma_{\rm QCD} = 1/\LDC^2$. 

\section{Conclusions}\label{concl}
We have reconsidered the relic  abundance of neutral colored relics, finding that
hadron collisions at temperatures below the QCD scale reduce it by a few orders of magnitude.
In particular we considered a quasi-stable gluino:
fig.\fig{Omega} shows its relic abundance, and
fig.\fig{constraints} the cosmological constraints,
taking into account the new effect and new data.

Co-annihilations between gluinos and neutralino DM are similarly strongly affected by confinement,
provided that their mass difference is smaller than a few GeV, as shown in fig.\fig{GluinoCAN}a.

In section~\ref{triplet} we considered charged colored relics, considering in particular the case of a quasi-stable stop.  
In this case, confinement  gives a big contribution to co-annihilations with neutralinos
only if $\tilde{t}\tilde{t}\tilde{t}$ baryons decay into SM particles via $\tilde t\tilde t \to tt$ before that
stop decays to neutralinos.


\small

\subsubsection*{Acknowledgements}
C.G. thanks the CERN Theoretical Physics Department for hospitality during the completion of this work. 
J.S. thanks the University of Florence and the Florence INFN division where a large fraction of this work was completed. J.S. is grateful to the support by the CP$^3$-Origins centre. The CP$^3$-Origins centre is partially funded by the Danish National Research Foundation, grant number DNRF90.
This work was supported by the ERC grant NEO-NAT. We thank Martti Raidal and NICPB for allowing us to run our codes on their machines.
The authors support equal opportunities.
\small

\appendix


\section{Non-abelian bound states}
Production cross sections and decay widths of two-body bound states
due to perturbative non-abelian gauge interactions
have been given in~\cite{1702.01141}, for bound states with low angular momentum
$\ell$.
Following the same notations, 
in section~\ref{Gamma} we generalise the decay widths to any $\ell$.
Although not needed in this work, in section~\ref{sigma} we also
show the cross sections for formation of bound states with generic $\ell$.  
We consider emission of a single-vector $V^a$  in dipole approximation, 
such that the angular momenta of the initial and final states differ by  $\Delta \ell=\pm 1$.
We denote with $\alpha$ the non-abelian gauge coupling,
with $M_a$ the vector mass,
and with $M$ the
common mass of the two particles which form the bound state.

\subsection{Cross sections for bound state formation}\label{sigma}
Production of a bound state with angular momentum $\ell$
proceeds from initial states with angular momentum $\ell\pm1$:
$(\sigma^{n\ell}_{{\rm bsf}} v_{\rm rel})_a=(\sigma^{n}_{{\rm bsf}} v_{\rm rel})_a^{\ell-1\to \ell}+(\sigma^{n}_{{\rm bsf}} v_{\rm rel})_a^{\ell+1\to \ell}$.
The cross sections are
\be
\begin{aligned}
(\sigma^{n}_{{\rm bsf}} v_{\rm rel})_a^{\ell+1\to \ell}&=\frac{8 (\ell+1)}{2\ell+3}\frac{\alpha k}{M^2}\left(1-\frac{k^2}{3\omega^2}\right) 
 \Bigg|\int r^2dr\times R^*_{n \ell ,j'i'} \\
&\times\Bigg(\frac{1}{2}\left(T_{i'i}^a\delta_{jj'}-\overline{T}_{j'j}^{a{*}}\delta_{ii'}\right)\left(\partial_r+\frac{\ell+2}{r}\right)
-i\,\frac{\alpha M}{2} \left(T_{i'i}^b\overline{T}_{j'j}^cf^{abc}\right) e^{-M_ar}\Bigg)R_{p,\ell+1,ij}\Bigg|^2
\end{aligned}
\ee
or, equivalently, integrating by parts
\be
\begin{aligned}
(\sigma^{n}_{{\rm bsf}} v_{\rm rel})_a^{\ell+1\to \ell}&=\frac{8 (\ell+1)}{2\ell+3}\frac{\alpha k}{M^2}\left(1-\frac{k^2}{3\omega^2}\right) 
 \Bigg|\int r^2drR_{p,\ell+1,ij}\times\\
&\times\Bigg(\frac{1}{2}\left(T_{i'i}^a\delta_{jj'}-\overline{T}_{j'j}^{a{*}}\delta_{ii'}\right)\left(\partial_r-\frac{\ell}{r}\right)
+i\,\frac{\alpha M}{2} \left(T_{i'i}^b\overline{T}_{j'j}^cf^{abc}\right) e^{-M_ar}\Bigg)R^*_{n\ell,j'i'}\Bigg|^2,
\end{aligned}
\ee
where $R_{n\ell,ij}$ is the bound state wave-function
in the two-particle space $|i\rangle \otimes | j\rangle$,
and $R_{p\ell,ij}$ is the wave-function of the
initial free state  with relative momentum $p$ and angular momentum $\ell$.
The other cross section is
\be
\begin{aligned}
(\sigma^{n}_{{\rm bsf}} v_{\rm rel})_a^{\ell-1\to \ell}&=  \frac{8 \ell}{2\ell -1}\frac{\alpha k}{M^2}\left(1-\frac{k^2}{3\omega^2}\right) 
 \Bigg|\int r^2drR_{n\ell,j'i'}^*\times\\
&\times\Bigg(\frac{1}{2}\left(T_{i'i}^a\delta_{jj'}-\overline{T}_{j'j}^{a{*}}\delta_{ii'}\right)\left(\partial_r-\frac{\ell-1}{r}\right)
-i\,\frac{\alpha M}{2} \left(T_{i'i}^b\overline{T}_{j'j}^cf^{abc}\right) e^{-M_ar}\Bigg)R_{p,\ell-1,ij}\Bigg|^2.
\end{aligned}
\ee
The formul\ae{} above simplify if the gauge group is unbroken,
or at least if all vectors have a common mass.
Then, a decomposition into irreducible representations 
allows to reduce the cross sections to abelian-like expressions:
\begin{eqnsystem}{sys:sigmasgen}
(\sigma^{n}_{{\rm bsf}} v_{\rm rel})_a^{\ell+1\to \ell }&=&\frac{8 (\ell+1)}{2 \ell +3}\frac{\alpha k}{M^2}\left(1-\frac{k^2}{3\omega^2}\right)\times \nonumber \\
&& \times \left|\int r^2dr R^*_{n \ell}\left(C_{\cal J}^{aMM'}\left(\partial_r+\frac{\ell+2}{r}\right)+C_{\cal T}^{aMM'}\frac{\alpha M}{2}e^{-M_a r}\right)R_{p, \ell +1}\right|^2 \nonumber 
\\
(\sigma^{n}_{{\rm bsf}} v_{\rm rel})_a^{\ell-1\to \ell}&=&\frac{8 \ell}{2\ell-1}\frac{\alpha k}{M^2}\left(1-\frac{k^2}{3\omega^2}\right)\times \nonumber \\
&&\times \left|\int r^2dr R^*_{n\ell}\left(C_{\cal J}^{aMM'} \left(\partial_r-\frac{\ell-1}{r}\right) + C_{\cal T}^{aMM'}\frac{\alpha M}{2}  e^{-M_ar}\right)R_{p, \ell-1}\right|^2 \nonumber
\end{eqnsystem}
where the group-theory part has been factored out in the coefficients
\begin{eqnsystem}{sys:CJT}
C_{\cal J}^{aMM'} &\equiv &\frac{1}{2}\,\hbox{CG}^M_{ij}\hbox{CG}^{M'*}_{i'j'}
 (T_{i'i}^a\delta_{jj'}+T_{j'j}^{a*}\delta_{ii'})
 =\frac{1}{2} \Tr[\hbox{CG}^{M'}\{\hbox{CG}^M,T^a\}]
 \\
C_{\cal T}^{aMM'} &\equiv& i\,\hbox{CG}^M_{ij}\hbox{CG}^{M'*}_{i'j'}  (T_{i'i}^bT_{jj'}^cf^{abc})=
i\Tr\Big[\hbox{CG}^{M'}\, T^b\,\hbox{CG}^M\,T^c\Big]f^{abc} \label{eq:nonabel}
\end{eqnsystem}
that holds separately for each initial channel $J$ and  final channel $J'$,
using the notations of~\cite{1702.01141}.

\subsection{Bound state decays}\label{Gamma}
The decay widths of a bound state trough single-vector emission
are obtained from the previous expressions
substituting the free-particle final state wave function $R_{p\ell}$
with the wave-function of the desired final bound states. 
Assuming again degenerate (or massless) vectors 
and a bound state in a representation $R$ with dimension $d_R$,
we find
\be
\begin{aligned}
\Gamma(n,\ell \to n', \ell-1)&=\frac 1 {d_R} \frac{8 \ell}{(2\ell+1)}\frac{\alpha k}{M^2}\left(1-\frac{k^2}{3\omega^2}\right) 
\times\\
&\times\sum_{aMM'}\left|\int r^2dr R^*_{n' ,\ell-1}\left(C_{\cal J}^{aMM'} \left(\partial_r+\frac{\ell+1}{r}\right)+C_{\cal T}^{aMM'}\frac{\alpha M}{2}  e^{-M_ar}\right)R_{n\ell}\right|^2
\end{aligned}
\ee
and
\be
\begin{aligned}
\Gamma(n,\ell \to n'', \ell+1)&=\frac 1 {d_R} \frac{8 (\ell+1)}{(2\ell+1)}\frac{\alpha k}{M^2}\left(1-\frac{k^2}{3\omega^2}\right) 
\times\\
&\times\sum_{aMM'}\left|\int r^2dr R^*_{n'' \ell+1}\left(C_{\cal J}^{aMM'} \left(\partial_r-\frac{l}{r}\right)+C_{\cal T}^{aMM'}\frac{\alpha M}{2}  e^{-M_ar}\right)R_{n\ell}\right|^2.
\end{aligned}
\ee



\footnotesize
\bibliographystyle{abbrv}

\begin{thebibliography}{nnn}\bibitem{hep-ph/9811386}
\article[hep-ph/9811386]{L. Giusti, A. Romanino, A. Strumia}{Nucl. Phys.}{B550}{3}{1998}
{Natural ranges of supersymmetric signals}.


\bibitem{1101.2195}
\article[1101.2195]{A. Strumia}{JHEP}{1104}{073}{2011}
{The Fine-tuning price of the early LHC}.


\bibitem{hep-th/0405159}
\article[hep-th/0405159]{N. Arkani-Hamed, S. Dimopoulos}{JHEP}{0506}{073}{2004}
{Supersymmetric unification without low energy supersymmetry and signatures for fine-tuning at the LHC}.


\bibitem{hep-ph/0409232}
\article[hep-ph/0409232]{N. Arkani-Hamed, S. Dimopoulos, G.F. Giudice, A. Romanino}{Nucl. Phys.}{B709}{3}{2004}
{Aspects of split supersymmetry}.


\bibitem{1108.6077}
\article[1108.6077]{G.F. Giudice, A. Strumia}{Nucl. Phys.}{B858}{63}{2012}
{Probing High-Scale and Split Supersymmetry with Higgs Mass Measurements}.


\bibitem{1407.4081}
\article[1407.4081]{E. Bagnaschi, G.F. Giudice, P. Slavich, A. Strumia}{JHEP}{1409}{092}{2014}
{Higgs Mass and Unnatural Supersymmetry}.


\bibitem{Mina} 
\article[hep-ph/0504210]{A. Arvanitaki, C. Davis, P.W. Graham, A. Pierce, J.G. Wacker}{Phys. Rev.}{D72}{075011}{2005}
{Limits on split supersymmetry from gluino cosmology}.


\bibitem{1801.01135}
\article[1801.01135]{V. De Luca, A. Mitridate, M. Redi, J. Smirnov, A. Strumia}{Phys. Rev.}{D97}{115024}{2018}
{Colored Dark Matter}.


\bibitem{hep-ph/0611322}
\article[hep-ph/0611322]{J. Kang, M.A. Luty, S. Nasri}{JHEP}{0809}{086}{2006}
{The Relic abundance of long-lived heavy colored particles}.


\bibitem{Toharia:2005gm}
\article[hep-ph/0503175]{M. Toharia, J.D. Wells}{JHEP}{0602}{015}{2005}
{Gluino decays with heavier scalar superpartners}.


\bibitem{Gambino:2005eh}
\article[hep-ph/0506214]{P. Gambino, G.F. Giudice, P. Slavich}{Nucl. Phys.}{B726}{35}{2005}
{Gluino decays in split supersymmetry}.


\bibitem{1402.6287}
\article[1402.6287]{A. De Simone, G.F. Giudice, A. Strumia}{JHEP}{1406}{081}{2014}
{Benchmarks for Dark Matter Searches at the LHC}.


\bibitem{1702.01141}
\article[1702.01141]{A. Mitridate, M. Redi, J. Smirnov, A. Strumia}{JCAP}{1705}{006}{2017}
{Cosmological Implications of Dark Matter Bound States}.


\bibitem{Griest:1989wd}
\article[Griest:1989wd]{K. Griest, M. Kamionkowski}{Phys. Rev. Lett.}{64}{615}{1990}
{Unitarity Limits on the Mass and Radius of Dark Matter Particles}.


\bibitem{1802.07720}
\article[1802.07720]{M. Geller, S. Iwamoto, G. Lee, Y. Shadmi, O. Telem}{JHEP}{1806}{135}{2018}
{Dark quarkonium formation in the early universe}.


\bibitem{QCDpert}
\article[Fischler:1977yf]{W. Fischler}{Nucl. Phys.}{B129}{157}{1977}
{$q$-$\bar q$ Potential in QCD}.
\article[hep-ph/9812205]{Y. Schroder}{Phys. Lett.}{B447}{321}{1998}
{The Static potential in QCD to two loops}.


\bibitem{lattice}
\article[1003.0936]{P. Bicudo}{Phys. Rev.}{D82}{034507}{2010}
{The QCD string tension curve, the ferromagnetic magnetization, and the quark-antiquark confining potential at finite Temperature}.
\article[1203.5320]{P. Petreczky}{J. Phys.}{G39}{093002}{2012}
{Lattice QCD at non-zero temperature}.
\article[1607.00299]{S. Aoki et al.}{Eur. Phys. J.}{C77}{112}{2017}
{Review of lattice results concerning low-energy particle physics}.
\article[hep-lat/0006022]{H.S. Bali}{JHEP}{D62}{114503}{2000}
{Casimir scaling of $\SU(3)$ static potentials}.


\bibitem{Hall:1984wk}
\article[Hall:1984wk]{R.L. Hall}{Phys. Rev.}{D30}{433}{1984}
{Simple eigenvalue formula for the Coulomb plus linear potential}.


\bibitem{Bhanot:1979vb}
\article[Bhanot:1979vb]{G. Bhanot, M.E. Peskin}{Nucl. Phys.}{B156}{391}{1979}
{Short Distance Analysis for Heavy Quark Systems. 2. Applications}.


\bibitem{1707.05380}
\article[1707.05380]{A. Mitridate, M. Redi, J. Smirnov, A. Strumia}{JHEP}{1710}{210}{2017}
{Dark Matter as a weakly coupled Dark Baryon}.


\bibitem{Kawasaki:2004qu}
\article[Kawasaki:2004qu]{M. Kawasaki, K. Kohri, T. Moroi}{Phys. Rev.}{D71}{083502}{2004}
{Big-Bang nucleosynthesis and hadronic decay of long-lived massive particles}.


\bibitem{Kawasaki:2017bqm}
\article[1709.01211]{M. Kawasaki, K. Kohri, T. Moroi, Y. Takaesu}{Phys. Rev.}{D97}{023502}{2018}
{Revisiting Big-Bang Nucleosynthesis Constraints on Long-Lived Decaying Particles}.


\bibitem{Pospelov:2006sc}
\article[Pospelov:2006sc]{M. Pospelov}{Phys. Rev. Lett.}{98}{231301}{2007}
{Particle physics catalysis of thermal Big Bang Nucleosynthesis}.


\bibitem{Kusakabe:2009jt}
\article[0906.3516]{M. Kusakabe, T. Kajino, T. Yoshida, G.J. Mathews}{Phys. Rev.}{D80}{103501}{2009}
{Effect of Long-lived Strongly Interacting Relic Particles on Big Bang Nucleosynthesis}.


\bibitem{Kawasaki:1994sc}
\article[Kawasaki:1994sc]{M. Kawasaki, T. Moroi}{Astrophys. J.}{452}{506}{1995}
{Electromagnetic cascade in the early universe and its application to the big bang nucleosynthesis}.


\bibitem{Wright:1993re}
\article[Wright:1993re]{E.L. Wirght et al.}{Astrophys. J.}{420}{450}{1994}
{Interpretation of the COBE FIRAS spectrum}.


\bibitem{Hu:1993gc}
\article[Hu:1993gc]{W. Hu, J. Silk}{Phys. Rev. Lett.}{70}{2661}{1993}
{Thermalization constraints and spectral distortions for massive unstable relic particles}.


\bibitem{Mather:1993ij}
\article[Mather:1993ij]{{\sc Firas} Collaboration}{Astrophys. J.}{420}{439}{1993}
{Measurement of the Cosmic Microwave Background spectrum by the COBE FIRAS instrument}.


\bibitem{Fixsen:1996nj}
\article[Fixsen:1996nj]{D.J. Fixsen, E.S. Cheng, J.M. Gales, J.C. Mather, R.A. Shafer, E.L. Wright}{Astrophys. J.}{473}{576}{1996}
{The Cosmic Microwave Background spectrum from the full COBE FIRAS data set}.


  

  


\bibitem{Cirelli:2010xx}
\article[1012.4515]{M. Cirelli, G. Corcella, A. Hektor, G. Hutsi, M. Kadastik, P. Panci, M. Raidal, F. Sala, A. Strumia}{JCAP}{1103}{051}{2010}
{PPPC 4 DM ID: A Poor Particle Physicist Cookbook for Dark Matter Indirect Detection}.


\bibitem{Kogut:2011xw}
\article[1105.2044]{{\sc PIXIE} Collaboration}{JCAP}{1107}{025}{2011}
{The Primordial Inflation Explorer (PIXIE): A Nulling Polarimeter for Cosmic Microwave Background Observations}.


\bibitem{Slatyer:2016qyl}
\article[1610.06933]{T.R. Slatyer, C-L. Wu}{Phys. Rev.}{D95}{023010}{2017}
{General Constraints on Dark Matter Decay from the Cosmic Microwave Background}.


\bibitem{Poulin:2016anj}
\article[1610.10051]{V. Poulin, J. Lesgourgues, P.D. Serpico}{JCAP}{1703}{043}{2017}
{Cosmological constraints on exotic injection of electromagnetic energy}.




\bibitem{Bowman:2018yin}
\article[Bowman:2018yin]{J.D. Bowman, A.E.E. Rogers, R.A. Monsalve, T.J. Mozdzen, N. Mahesh}{Nature}{555}{67}{2018}
{An absorption profile centred at 78 megahertz in the sky-averaged spectrum}.


\bibitem{1803.09739}
\article[1803.09739]{H. Liu, T.R. Slatyer}{Phys. Rev.}{D98}{023501}{2018}
{Implications of a 21-cm signal for dark matter annihilation and decay}.


\bibitem{1803.09390}
\article[1803.09390]{S. Clark, B. Dutta, Y. Gao, Y-Z. Ma, L.E. Strigari}{Phys. Rev.}{D98}{043006}{2018}
{21 cm limits on decaying dark matter and primordial black holes}.


\bibitem{1803.11169}
\article[1803.11169]{A. Mitridate, A. Podo}{JCAP}{1805}{069}{2018}
{Bounds on Dark Matter decay from 21 cm line}.


\bibitem{Cohen:2016uyg}
\article[1612.05638]{T. Cohen, K. Murase, N.L. Rodd, B.R. Safdi, Y. Soreq}{Phys. Rev. Lett.}{119}{021102}{2017}
{$\gamma$-ray Constraints on Decaying Dark Matter and Implications for IceCube}.


\bibitem{Foster:1998wu}
\article[Foster:1998wu]{M. Foster, C. Michael}{Phys. Rev.}{D59}{094509}{1999}
{Hadrons with a heavy color adjoint particle}.


\bibitem{1011.2964}
\article[1011.2964]{G.R. Farrar, R. Mackeprang, D. Milstead, J.P. Roberts}{JHEP}{1102}{018}{2010}
{Limit on the mass of a long-lived or stable gluino}.


\bibitem{Polikanov:1990sf}
\article[Polikanov:1990sf]{S. Polikanov, C.S. Sastri, G. Herrmann, K. Lutzenkirchen, M. Overbeck, N. Trautmann, A. Breskin, R. Chechik, Z. Frankel}{Z. Phys.}{A338}{357}{1991}
{Search for supermassive nuclei in nature}.


\bibitem{Ellis:2015vaa}
\article[1503.07142]{J. Ellis, F. Luo, K.A. Olive}{JHEP}{1509}{127}{2015}
{Gluino Coannihilation Revisited}.


\bibitem{ATLAS:2018yey}
\heparticle[ATLAS:2018yey]{ATLAS Collaboration}{Reinterpretation of searches for supersymmetry in models with variable $R$-parity-violating coupling strength and long-lived $R$-hadrons}


\bibitem{di-jet} 
\article[1703.09127]{{\sc ATLAS} Collaboration}{Phys. Rev.}{D96}{052004}{2017}
{Search for new phenomena in dijet events using 37 fb$^{-1}$ of $pp$ collision data collected at $\sqrt{s}=$13 TeV with the ATLAS detector}


\bibitem{1402.0492}
\article[1402.0492]{M. Papucci, K. Sakurai, A. Weiler, L. Zeune}{Eur. Phys. J.}{C74}{3163}{2014}
{Fastlim: a fast LHC limit calculator}


\bibitem{hep-ph/0607290}
\article[hep-ph/0607290]{Y. Jia}{JHEP}{0610}{073}{2006}
{Variational study of weakly coupled triply heavy baryons}.


\bibitem{1008.3154}
\article[1008.3154]{S. Meinel}{Phys. Rev.}{D82}{114514}{2010}
{Prediction of the $\Omega_{bbb}$ mass from lattice QCD}.


\bibitem{1212.6847}
\article[1212.6847]{A. Delgado, G.F. Giudice, G. Isidori, M. Pierini, A. Strumia}{Eur. Phys. J.}{C73}{2370}{2013}
{The light stop window}.


\bibitem{1011.1054}
\article[1011.1054]{M. Pospelov, J. Pradler}{Ann. Rev. Nucl. Part. Sci.}{60}{539}{2010}
{Big Bang Nucleosynthesis as a Probe of New Physics}.




\end{thebibliography}

   
\end{document}